%% file: main.tex
\documentclass[a4paper,11pt,openany]{book}
\usepackage[centertags]{amsmath}
\usepackage{amscd}
\usepackage{amsthm}
\usepackage{amssymb}
\usepackage{enumerate}
\usepackage{multicol}
\usepackage{url}

\usepackage[english,catalan,spanish]{babel}
\usepackage{color}
\usepackage{tikz}
\usepackage{tabls} 
\usepackage{indentfirst} 
\usepackage{float}    

\usepackage[T1]{fontenc}
\usepackage[sc]{mathpazo}
\usepackage[latin1]{inputenc}
\usepackage[twoside,bindingoffset=1cm]{geometry}

\usepackage{subfiles}
\usepackage{algorithm}
\usepackage{algpseudocode}

\usepackage{subfig}

\newtheorem{defn0}{Definition}[chapter]
\newtheorem{prop0}[defn0]{Proposition}
\newtheorem{thm0}[defn0]{Theorem}
\newtheorem{lemma0}[defn0]{Lemma}
\newtheorem{corollary0}[defn0]{Corollary}
\newtheorem{example0}[defn0]{Example}
\newtheorem{remark0}[defn0]{Remark}
\newtheorem{conjecture0}[defn0]{Conjecture}
\newtheorem*{post}{Postulate}

\newenvironment{definition}{ \begin{defn0}}{\end{defn0}}

\newenvironment{theorem}{\bigskip \begin{thm0}}{\end{thm0}}

\usepackage{fancyhdr}
\newif\ifprivate
\privatetrue

\def\???{\ifprivate {\bf {???}} \marginpar{\begin{center}{\Huge {\bf ?}}\end{center}}
\else \fi}
\newcommand{\m}{\mathfrak m}

\begin{document}

\pagestyle{empty}

\begin{titlepage}
\begin{center}
\begin{figure}[htb]
\begin{center}
\includegraphics[width=6cm]{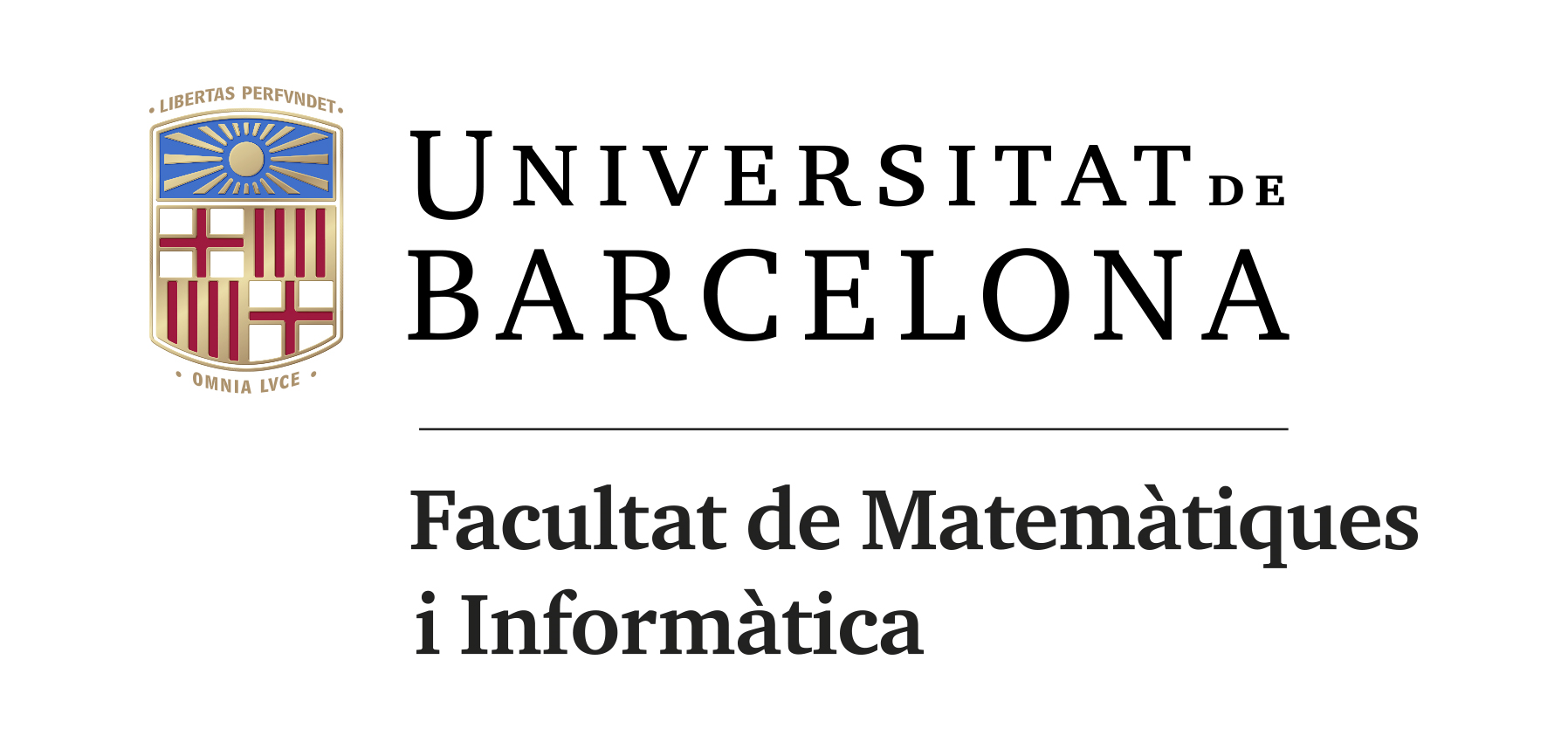}
\end{center}
\end{figure}

\vspace*{0.5cm}
\textbf{\large BACHELOR'S DEGREE IN MATHEMATICS \\ BACHELOR'S DEGREE IN COMPUTER SCIENCE AND SOFTWARE ENGINEERING} \\
\vspace*{.5cm}
\textbf{\LARGE Bachelor's Final Thesis} \\

\vspace*{1cm}
\rule{16cm}{0.1mm}\\
\begin{Huge}
\textbf{QUANTUM ANNEALING AND TENSOR NETWORKS: A POWERFUL COMBINATION TO SOLVE OPTIMIZATION PROBLEMS} \\
\end{Huge}
\rule{16cm}{0.1mm}\\

\vspace{0.5cm}

\begin{flushright}
\textbf{\LARGE Author: Miquel Albert\'{i} Binimelis}

\vspace*{0.5cm}

\renewcommand{\arraystretch}{1.5}
\begin{tabular}{ll}
\textbf{\Large Directors:} & \textbf{\Large Dr. Nahuel Norbeto Statuto } \\
& \textbf{\Large Dr. Luis V\'{i}ctor Dieulefait} \\
\textbf{\Large Conducted at:} & \textbf{\Large  Faculty of Mathematics and Computer Science, } \\
& \textbf{\Large University of Barcelona} \\
\\
\textbf{\Large Barcelona, June 2024}
\end{tabular}

\end{flushright}

\end{center}

\end{titlepage}

\newpage
\selectlanguage{english}
\pagenumbering{roman} \setcounter{page}{0}
\tableofcontents
\newpage \thispagestyle{empty}

\pagestyle{fancy}

\section*{Abstract}

{\let\thefootnote\relax\footnote{2020 Mathematics Subject Classification. 15A18, 47B02, 68Q12, 68W01, 81T32, 90C20, 90C27}}

Quantum computing has long promised to revolutionize the way we solve complex problems. At the same time, tensor networks are widely used across various fields due to their computational efficiency and capacity to represent intricate systems. While both technologies can address similar problems, the primary aim of this thesis is not to compare them. Such comparison would be unfair, as quantum devices are still in an early stage, whereas tensor network algorithms represent the state-of-the-art in quantum simulation. Instead, we explore a potential synergy between these technologies, focusing on how two flagship algorithms from each paradigm, the Density Matrix Renormalization Group (DMRG) and quantum annealing, might collaborate in the future.

Furthermore, a significant challenge in the DMRG algorithm is identifying an appropriate tensor network representation for the quantum system under study. The representation commonly used is called Matrix Product Operator (MPO), and it is notoriously difficult to obtain for certain systems. This thesis outlines an approach to this problem using finite automata, which we apply to construct the MPO for our case study.

Finally, we present a practical application of this framework through the quadratic knapsack problem (QKP). Despite its apparent simplicity, the QKP is a fundamental problem in computer science with numerous practical applications. In addition to quantum annealing and the DMRG algorithm, we implement a dynamic programming approach to evaluate the quality of our results.

Our results highlight the power of tensor networks and the potential of quantum annealing for solving optimization problems. Moreover, this thesis is designed to be self-explanatory, ensuring that readers with a solid mathematical background can fully understand the content without prior knowledge of quantum mechanics.

\vspace{1.5cm}

{\bf Keywords:} Tensor networks, quantum computing, quantum annealing, Matrix Product Operators, DMRG algorithm, finite automata, QUBO formulation, quadratic knapsack problem, optimization, dynamic programming.

\mainmatter
\section{Introduction}

Quantum computers and Tensor Networks are engaged in an exciting battle to simulate and understand the intricate nature of quantum systems. The remarkable capacity of tensor networks to represent and manipulate the quantum realm is impressive \cite{TNBasicIntro, GlenEvenbly2022practical, Bridgeman_2017, Or_s_2014}. Algorithms such as the Density Matrix Renormalization Group (DMRG) have captured considerable attention \cite{DMRG, video_lecture_DMRG}, and will be one of the main focuses of this thesis. However, the implications of tensor networks extend beyond physics, as many industrial problems can be solved using these techniques when cast as optimization problems. 

On the other hand, we find ourselves in the Noisy Intermediate-Scale Quantum (NISQ) era \cite{9251243}, in which quantum computers still cannot outperform classical ones in real-world applications. Among the various types of quantum computers, quantum annealers exhibit faster scalability in terms of qubit count due to their relative ease of construction \cite{Albash_2018, combarro2023practical}. Although these devices are not universal quantum computers, they excel at solving specific optimization tasks formulated as Quadratic Unconstrained Binary Optimization (QUBO) problems.

In this project, the reader will obtain an overview of all of these concepts, with a particular emphasis on their synergies. Our technical contributions are:
\begin{itemize}
    \item The QUBO formulation of the quadratic knapsack problem with the unbalanced penalization technique \cite{montanezbarrera2023unbalanced}, that allows the encoding of inequality constraints without adding more variables.
    \item The creation of the Matrix Product Operator (MPO) that describes the quantum system used in quantum annealing: the annealing Hamiltonian of an Ising model. This enables the use of the DMRG algorithm to study the evolution performed in quantum annealing.
    \item A proposal of an annealing scheduling time evolution based on the study of the minimum gap with the DMRG.
\end{itemize}

\noindent Other valuable insights of the thesis are:
\begin{itemize}
    \item An introduction to quantum computing and tensor networks for people with no physics background.
    \item An explanation of how the DMRG and quantum annealing algorithms work.
    \item A detailed description of how to create MPOs using finite automata.
    \item A Python implementation of all these algorithms, along with notebooks demonstrating their use.
\end{itemize}

The thesis is organized into five chapters. The first two chapters introduce the basics of quantum computing and tensor networks, respectively. The third chapter integrates this knowledge to explore how tensor networks can be used to simulate quantum mechanics, and consequently quantum annealing, to study their performance and combine efforts for better results. The fourth chapter focuses on a methodology using finite automata to create MPOs, which describe a quantum system with tensor networks for use in DMRG. This methodology is then applied to derive the expression for the system of interest. Finally, chapter five applies all the concepts learned to tackle the quadratic knapsack problem using different algorithms: dynamic programming, DMRG, and simulated quantum annealing.

The main abbreviatures that will be used are:

\begin{itemize}
    \item QUBO: Quadratic Unconstrained Binary Optimization
    \item AQC: Adiabatic Quantum Computing
    \item QA: Quantum Annealing
    \item TN: Tensor Network
    \item SVD: Singular Value Decomposition
    \item MPS: Matrix Product State
    \item MPO: Matrix Product Operator
    \item DMRG: Density Matrix Renormalization Group
    \item QKP: Quadratic Knapsack Problem
    \item DP: Dynamic Programming
\end{itemize}

All the Python code can be accessed from the corresponding GitHub public repository\footnote{https://github.com/MiquelAlberti2/Quantum-annealing-and-tensor-networks}. An explanation of the structure of the code can be found on the README page.

\subfile{chapters/C1_mathematical_framework}

\subfile{chapters/C2_TN}

\subfile{chapters/C3_Connection_TN_Quantum}

\subfile{chapters/C4_MPO}

\subfile{chapters/C5_QKP}

\subfile{chapters/Conclusion}

\appendix

\chapter{First Appendix: other algorithms}

\section{The power method}
\label{sec: power method}

In the optimization step of the Density Matrix Renormalization Group (DMRG) algorithm, the objective is to find a good approximation of the eigenvector associated with the lowest eigenvalue of the Hamiltonian matrix. The algorithm typically used for this purpose is the Lanczos method, which is a more complex variant of the power method.

The power method is used to approximate a matrix's eigenvector. It involves repeatedly applying a matrix to a vector, which should eventually converge towards the direction of the eigenvector associated with the largest eigenvalue.

Let $A \in \mathbb{C}^{n \times n}$ be a diagonalizable matrix. Let's suppose its eigenvalues satisfy
$$
|\lambda_1| > |\lambda_2| \ge |\lambda_3| \ge \ldots \ge |\lambda_n| \; \; .
$$

The power method starts with an initial vector $z^{(0)} \in \mathbb{R}^n$ (could be a random one, as long as it is not null). Then we build a series of vectors
$$
z^{(k+1)} = A z^{(k)}, \; \; \; k=0,1,2,\ldots \; \; ,
$$
and its direction will converge to the direction of the eigenvector with eigenvalue $\lambda_1$. To show that, let's consider a basis of eigenvectors $\{ v_1, \ldots, v_n \}$ where $v_i$ has eigenvalue $\lambda_i$. Then, we can write $z^{(0)}$ as 
$$
z^{(0)} = \alpha_1 v_1 + \ldots + \alpha_n v_n \; \; .
$$
Therefore,
$$
z^{(k)} = A^k z^{(0)} = A^k \sum_i^n \alpha_i v_i = \sum_i^n \lambda_i^k \alpha_i v_i = \lambda_1^k \left[ \alpha_1 v_1 + \sum_{i=2}^n \left( \frac{\lambda_i}{\lambda_1}\right)^k \alpha_i v_i \right] \; \; .
$$
As $ \left( \lambda_i / \lambda_1 \right)^k \to 0 $ for $i>1$ when $k \to \infty$ then the direction of $z^{(k)}$ tends towards the direction of $v_1$ (if $\alpha_1 \neq 0$). In practice, we should also normalize the series because $\lambda_1^k$ could tend towards 0 or infinity.

However, in the DMRG algorithm, we are interested in the lowest eigenvector and eigenvalue. This can be achieved by applying the same algorithm to $A^{-1}$ (which exists in this case as hermitian operators are invertible). This is because the eigenvalues of $A^{-1}$ are $1 / \lambda_i$, so they satisfy
$$
\frac{1}{\left|\lambda_1\right|}<\frac{1}{\left|\lambda_2\right|} \leq \frac{1}{\left|\lambda_3\right|} \leq \ldots \leq \frac{1}{\left|\lambda_n\right|} .
$$
If $\lambda_n<\lambda_{n-1}$ we are under the same conditions as before.

The Lanczos method computes the lowest eigenvector and eigenvalue based on this idea. Here, $z^{(0)}$ would be the tensor $B_{12}$ of Fig. \ref{fig: B12_H_mult} reshaped into a vector, and the successive application of the matrix $A$ corresponds to the operation described in the same figure with the matrix $H$. However, the actual process is much more complex than this explanation, and it does not rely on the inverse matrix as computing it is computationally expensive. For a detailed description of the Lanczos method, refer to Ref. \cite{Lanczos}.

The primary reference for this appendix is my lecture notes from the university course \textit{Numerical Methods II} taught by Professor Joan Carles Tatjer at the University of Barcelona.

\section{Dynamic programming approach for the quadratic knapsack problem}
\label{sec: DP algorithm}

Dynamic programming (DP) is the most standard approach to obtain the exact solution for the knapsack problem (KP), with a time complexity of \(\text{\( O(N \cdot W) \)}\). This does not contradict the fact that the KP is an NP-hard problem since $W$ is not polynomial as a function of the input length.

However, the quadratic knapsack problem (QKP) is harder as it does not obey Bellman's principle of optimality \cite{dp_QKP}, which states that at any time in a multi-stage decision process (is item $i$ in the optimal solution or not), the remaining decisions (are items $i+1, \ldots, N$ in the optimal solution or not) must constitute an optimal policy with respect to the current decisions made (are items $1, \ldots, i-1$ in the optimal solution or not). Mathematically \cite{Moura2014}, 
$$
V_k(x_k) = max_{d_k}\{ g(x_k, d_k) + V_{k+1}(x_{k+1})\}, \; \; \; k=1, \ldots, N
$$
where $k$ indicates in which iteration we are, $x_k$ is the state at step $k$ (i.e. decisions made so far), $V_k(x_k)$ is the global optimal value being in step $k$ and state $x_k$, and $g(x_k, d_k)$ is the instantaneous cost of taking decision $d_k$ in the state $x_k$. This is the basis of DP as guarantees that we can get an optimal solution by solving the problem backward, starting from $k=N$ and using the computation of $V_{k+1}$ to find the optimal $d_k$ at each step.

Why does this principle not hold in QKP? In a classical KP, once we finish an iteration for item $i$, we know that the decisions of including or not the item for all the weight capacities up to $W$ are aligned with the global optimum solution \cite{dp_QKP}. However, in QKP we cannot make that assumption because we are not taking into account the potential extra profit that item $i$ could have when paired with items $i+1, \ldots N$.

Despite DP cannot give us the optimal solution for QKP, it can yield good approximations, especially for instances when the extra profit given for pairs is less relevant (as the problem gets closer to a classical KP).
\begin{algorithm}
\caption{DP approach for QKP}\label{alg: DP}
\begin{algorithmic}
\Require $v_{ij}$ to be a triangular matrix

\State Initialize $V(k) \gets 0$ for $k=1, \ldots W$ \Comment{stores optimal profit for each weight}
\State Initialize $S(k) \gets \emptyset $ for $k=1, \ldots W$ \Comment{stores solution items for each weight}

\For{$k = 1, \ldots, N$}
\For{$r = W, \ldots, 1$}

\If{$w_k < r$}
    \State $v \gets V(r-w_k) + v_{ii}$
    \For{$i \in S(r-w_k)$}
        \State $v \gets v + v_{max(i,k), min(i,k)}$
    \EndFor

    \If{$V(r) < v$}    \Comment{if item $k$ improves solution, update $V$ and $S$}
        \State $V(r) \gets v$
        \State $S(r) \gets S(r) \cup \{k\}$
    \EndIf
\EndIf
\EndFor
\EndFor
\State Output $V(W)$ and $S(W)$
\end{algorithmic}
\end{algorithm}

Time complexity is roughly $O(W\cdot N^{2})$. We say roughly because the nested loop over $N$ is just over the items included in the best solution so far. This is in fact the key difference concerning the algorithm for the classical KP: the need to keep track of the items in the solutions of the subproblems to compute the extra profit given by pairs of items. 

\chapter{Second Appendix: explicit representation of the MPO for the annealing Hamiltonian}
\label{sec: Tables MPO}

Here we present our implementation of the Matrix Product Operator (MPO) for the annealing Hamiltonian of an Ising model
\begin{equation*}
    H=-(1-s) \sum_i X_i+s\left(\sum_i h_i Z_i+\sum_{i j} J_{i j} Z_i Z_j\right) \; \; .
\end{equation*}
We will be using the notation introduced in Ref. \cite{AppendixA}. The following tables illustrate how to fill the matrices of the MPO. In brief, the (left, right)-input specifies the matrix's position, and the output indicates the operator to place there. For a more detailed explanation, refer to Chapter \ref{sec: MPO}.

\begin{center}
Rules for $M_k$, $\forall k$. \\
Let -1 and -2 be the last and the second last indices, respectively.
\end{center}
$$
\begin{array}{c|ccc}
\text { rule-number } & \text { (left, right)- input } & & \text { output } \\
\hline
1 & (1,1) & \rightarrow & \mathbb{1} \\
2 & (-1,-1) & \rightarrow & \mathbb{1} \\
3 & (-2,-1) & \rightarrow & Z \\
4 & (1,-1) & \rightarrow & - (1-s)X + s\cdot h_{k}\; Z
\end{array}
$$
\vspace{0.5cm}
\begin{center}
Additional rules for $M_k$, $k<\lfloor N/2 \rfloor$. The matrix dimension is $k+2 \times k+3$. \\
$m = 2, \ldots, k$.
\end{center}
$$
\begin{array}{c|ccc}
\text { rule-number } & \text { (left, right)- input } & & \text { output } \\
\hline
5 & (1,2) & \rightarrow & Z \\
6 & (1,k+2) & \rightarrow & s\cdot J_{k, k+1} \; Z \\
7 & (m, m+1) & \rightarrow & \mathbb{1} \\
8 & (m, k+2) & \rightarrow & s \cdot J_{k-m+1, k+1}\; \mathbb{1}
\end{array}
$$
\vspace{0.5cm}
\begin{center}
Additional rules for $M_{\lfloor N/2 \rfloor}$. The matrix dimension is $\lfloor N/2 \rfloor+2 \times \lfloor N/2 \rfloor + a$. \\
$m = 2, \ldots, k$; $n=2, \ldots, \lfloor N/2 \rfloor + a -1$ \\
$a=2$ if $N$ is even, $a=3$ otherwise.
\end{center}
$$
\begin{array}{c|ccc}
\text { rule-number } & \text { (left, right)- input } & & \text { output } \\
\hline
5 & (1,n) & \rightarrow & s\cdot J_{\lfloor N/2 \rfloor, N-n+2}\; Z \\
6 & (m,n) & \rightarrow & s\cdot J_{\lfloor N/2 \rfloor - m +1, N-n+2}\;  \mathbb{1}
\end{array}
$$
\vspace{0.5cm}
\begin{center}
Additional rules for $M_k$, $k>\lfloor N/2 \rfloor$. The matrix dimension is $N-k+3 \times N-k+2$. \\
$m = 2, \ldots, N-k+1$.
\end{center}
$$
\begin{array}{c|ccc}
\text { rule-number } & \text { (left, right)- input } & & \text { output } \\
\hline
5 & (1,m) & \rightarrow & s\cdot J_{k,N-m+2}\; Z \\
6 & (m,m) & \rightarrow & \mathbb{1}
\end{array}
$$

\vspace{0.5cm}
The implementation can be found in the corresponding GitHub public repository\footnote{https://github.com/MiquelAlberti2/Quantum-annealing-and-tensor-networks}. While we did not find an MPO for this Hamiltonian in the existing literature, we validated its correctness through two methods. First, for small instances, we contracted the MPO and confirmed it matched the symbolic expression in \texttt{Visualizations/symbolic\_annealing\_ham.ipynb}. For larger instances, we diagonalized both the real Hamiltonian matrix and the matrix obtained from contracting the MPO (see \texttt{solutions\_enhaced\_annealing.ipynb}). Both matrices had the same spectral decomposition, particularly the same ground state.

\bibliographystyle{amsplain}
\nocite{*}
\bibliography{refs}

\end{document}

%% file: chapters/C1_mathematical_framework.tex
\chapter{Mathematical framework of quantum mechanics and quantum computing}

When we talk about classical computing, we know that boolean algebra is the mathematical framework that perfectly describes the structure and patterns to manipulate the minimum unit of information: bits. However, with quantum computing, we need a framework that captures the complexity of quantum mechanics, and it turns out that linear algebra does just that in a surprisingly natural way. In this section, we will explain how linear algebra is used in this context so that people with no physics background can follow this thesis simply by understanding the maths.

The minimum unit of information in quantum computing is a qubit (quantum bit), which encapsulates the state of a two-level quantum system. By two-level, we mean it has two distinguishable elemental states, denoted as $|0\rangle$ and $|1\rangle$. However, unlike classical bits (which also have two distinguishable states), qubits can be in a mixture of these two states, a phenomenon known as superposition \cite{nielsen00}. Mathematically, a superposition is just a complex linear combination:
$$
|\psi\rangle=\alpha_0|0\rangle+\alpha_1|1\rangle=\binom{\alpha_0}{\alpha_1} \; ; \;  \alpha_0, \alpha_1 \in \mathbb{C} , \; \left|\alpha_0\right|^2+\left|\alpha_1\right|^2=1,
$$
where $|\alpha|^2=\alpha \cdot \bar{\alpha}$ denotes the modulus of $\alpha$. The coefficients $\alpha_0$ and $\alpha_1$, called amplitudes, reflect how close the state $|\psi\rangle$ is to the elemental states $|0\rangle$ or $|1\rangle$, and they are normalized because it is, indeed, a probability distribution. We can manipulate these superposition states, but once we measure them they have to collapse to an elemental state, with a probability given by the amplitudes: we will measure the state $|0\rangle$ with probability $\left|\alpha_0\right|^2$ or the state $|1\rangle$ with probability $\left|\alpha_1\right|^2$.

\section{Hilbert space}
\label{sec: hilbert space}

We have already introduced the symbol $| \; \rangle$. This is known as a \textit{ket} and is just the standard notation to designate a vector representing a quantum state (a state vector). In a one-qubit system, we have the basis
\begin{multicols}{2}
\noindent
$$
|0\rangle=\binom{1}{0}
$$
\columnbreak
$$
|1\rangle=\binom{0}{1} \; \; .
$$
\end{multicols}
\noindent but we can have systems with many more particles (and therefore states). The set of all possible state vectors describing a given physical system forms a complex vector space.

The most convenient basis when manipulating qubits is the one given by combining the previous basis with the kronecher product (or tensor product) denoted by $\otimes$. For a system with $N$ qubits it is
$$
\{ |i_1, \ldots, i_N\rangle := |i_1\rangle \otimes \ldots \otimes |i_N\rangle \; | \; i_1, \ldots i_N \in \{0,1\} \}.
$$
For example, for a system with 2 qubits this is

\begin{multicols}{2}
\noindent
\begin{equation*}
|00\rangle := |0\rangle \otimes |0\rangle = \binom{1}{0} \otimes \binom{1}{0} = \begin{pmatrix} 1 \\ 0 \\ 0  \\ 0 \end{pmatrix}
\end{equation*}

\vspace{0.3cm}
\begin{equation*}
|01\rangle = \binom{1}{0} \otimes \binom{0}{1} = \begin{pmatrix} 0 \\ 1 \\ 0  \\ 0 \end{pmatrix}
\end{equation*}

\columnbreak

\noindent
\begin{equation*}
|10\rangle  = \binom{0}{1} \otimes \binom{1}{0} = \begin{pmatrix} 0 \\ 0 \\ 1  \\ 0 \end{pmatrix}
\end{equation*}

\vspace{0.3cm}
\begin{equation*}
|11\rangle = \binom{0}{1} \otimes \binom{0}{1} = \begin{pmatrix} 0 \\ 0 \\ 0  \\ 1 \end{pmatrix}.
\end{equation*}

\end{multicols}

With this example, we can clearly see the dimension of this vector space. It is $2^N$ where $N$ is the number of qubits. It is massive, as it grows exponentially with the number of qubits, thereby constituting the inherent potential of quantum computing.

On this vector space, we can define a metric. For that, we introduce the \textit{bra} vector $\langle \psi |$, which is the dual vector of $| \psi \rangle$, i.e. the one we obtain by doing the complex conjugate of the elements in the transpose vector. Given a vector space with basis $\{ | \psi_1 \rangle \ldots | \psi_n \rangle \}$ the ket and corresponding bra vector are defined as

\begin{align*}
    &| \psi \rangle = \sum_i^n \alpha_i | \psi_i \rangle && \langle \psi | = \sum_i^n \alpha_i^* \langle \psi_i | \; \; ,
\end{align*}

where $\alpha_i^*$ is the complex conjugate of $\alpha_i$. We can think of ket vectors as column vectors and bra vectors as row vectors.

Now, we define the inner product of $| \psi \rangle, | \phi \rangle$ as the vector product of the bra vector $\langle \psi |$ with the ket vector $| \phi \rangle$, denoted for simplicity as $\langle \psi | \phi \rangle$. As any inner product, it can be understood as the overlap of these two state vectors. However, having the particularity that now coordinates determines the probabilities of measuring a certain basis state: if $| \psi \rangle$ is an elemental state, the likelihood of observing the system to be in $| \psi \rangle$ given that it is in the state $| \phi \rangle$ will be given by $| \langle \psi | \phi \rangle | ^2$. This indeed denotes a probability distribution \cite{nielsen00} if the norm of the vector $\|| \phi \rangle\| := \sqrt{\langle \phi | \phi \rangle}$ is equal to 1. Otherwise, the state would not be physically valid. This will become clearer in Section \ref{sec: operators} when we discuss projection operators, as the inner product of vectors with norm 1 can be seen as the projection.

This complex vector space equipped with the inner product defined forms a Hilbert space $\mathcal{H}$ \cite{berberian1999introduction, nielsen00}.

\begin{definition}
    We say that a vector space equipped with an inner product is a Hilbert space $\mathcal{H}$ if it is complete, i.e. if every Cauchy sequence converges.
\end{definition}

We will not delve further into this definition and its implications, as it is not necessary for this project.

\subsubsection{Entanglement}

We used the Kronecker product to define the basis of an N-qubit system from the basis of each single-qubit system. In general, given any two Hilbert spaces $\mathcal{H}_1$ and $\mathcal{H}_2$ with basis $\{|\phi_i \rangle\}_{i=1, \ldots, n}$ and $\{|\psi_j \rangle\}_{j=1, \ldots, m}$, the tensor product $\mathcal{H}_1 \otimes \mathcal{H}_1$ is another Hilbert space with basis $\{|\phi_i \rangle\ \otimes |\psi_j \rangle\}_{i=1, \ldots, n; \; j=1, \ldots, m}$. We can use this tensor product to ``combine'' states in different Hilbert spaces \cite{lecture_delft, nielsen00}.

\begin{definition}
    Given any two Hilbert spaces $\mathcal{H}_1$ and $\mathcal{H}_2$ and a state in each of them $|\phi_1 \rangle \in \mathcal{H}_1$, $|\phi_2 \rangle \in \mathcal{H}_2$, we define their joint state as $|\phi_1 \phi_2 \rangle := |\phi_1 \rangle \otimes |\phi_2 \rangle \in \mathcal{H}_1 \otimes \mathcal{H}_1$.
\end{definition}
However, given a joint state $|\phi \rangle$, can we always decompose it as the tensor product of lower-dimensional Hilbert spaces? The answer in general is no, and that's because of the phenomenon called \textit{entanglement}.

To illustrate this, let's consider a state $|\phi \rangle$ in a 2-qubit system, $|\phi \rangle = (\alpha_0, \alpha_1, \alpha_2,\alpha_3)^T$. If we want to express $|\phi \rangle$ as the tensor product of two 1-qubit states $|\phi \rangle = |\phi_1 \phi_2 \rangle$ we would need to solve the system of equations:

$$
\binom{a_0}{a_1} \otimes\binom{b_0}{b_1}=\left(\begin{array}{l}
\alpha_0 \\
\alpha_1 \\
\alpha_2 \\
\alpha_3
\end{array}\right) \Longleftrightarrow\left\{\begin{array}{l}
a_0 b_0=\alpha_0 \\
a_0 b_1=\alpha_1 \\
a_1 b_0=\alpha_2 \\
a_1 b_1=\alpha_3
\end{array}\right.
$$

Which does not always have a solution. For instance, let's consider $|\phi \rangle = \left(\frac{1}{\sqrt{2}}, 0, 0,\frac{1}{\sqrt{2}}\right)^T$, which is a valid state because it has norm 1. Notice that 
$$
\alpha_1 = \alpha_2 = 0 \implies \left( a_0=0 \lor b_1=0 \right) \land \left( a_1=0 \lor b_0=0 \right) \; \; .
$$
Now, if $a_0=0$ then $a_1=1$ to ensure that $\langle \phi_1 |\phi_1 \rangle = 1$. But $a_1=1 \implies b_0=0 \implies \alpha_0 = 0$ which is a contradiction. The same happens with $\alpha_3$ if $b_1 = 0$. 

Let's give a physical explanation of what has happened. The fact that $\alpha_1 = \alpha_2 = 0$ but $\alpha_0 = \alpha_3 \neq 0$ implies that the only observable states when measuring $|\phi \rangle$ are $|00 \rangle$ or $|11 \rangle$ (with probability 1/2 each). This means these two particles cannot be described independently: if we measure one, the other would collapse into the same state. In this case, we say that particles are entangled, which is a key property when studying quantum computing algorithms.

On the other hand, we say that a quantum system is in a \textit{product state} if it can be written as a tensor product of the states of its components.

\section{Operators}
\label{sec: operators}

To exploit the properties of quantum states to do computation we need to manipulate them. By manipulation, we mean a physical process that transforms one state into another (of the same Hilbert space), that carries some useful information. Mathematically, a state is nothing more than a vector, hence we can represent these transformations by matrices. However, the resulting state after any manipulation needs to be a valid physical state, i.e. the mapping needs to preserve the (unitary) norm of the state.

\subsubsection{Unitary and hermitian operators}
\label{sec: unitary and hermitian operators}

There are many types of operators, but the second postulate of quantum mechanics \cite{nielsen00} tells us which are the ones we should care about.

\begin{post}
The time evolution of a closed quantum system is described by a unitary transformation. That is, the state $| \psi \rangle$ of the system at time $t_1$ is related to the state $| \psi' \rangle$ of the system at time $t_2$ by a \textbf{unitary operator} $U$ which depends only on the times $t_1$ and $t_2$,
$$
| \psi' \rangle = U| \psi \rangle \; \; .
$$
\end{post}

Unitary operators are the ones described by unitary matrices.

\begin{definition}  \label{unitary matrix}
In linear algebra, an invertible, complex, square matrix $U$ is \textit{unitary} if $U^{-1} = U^{*}$, where $U^{*}$ is its conjugate transpose. In physics, the conjugate transpose is usually referred to as the \textit{Hermitean adjoint}, denoted by a dagger $U^{\dagger}$.
\end{definition}

Furthermore, it is easy to observe that the Hermitian adjoint satisfies

$$
U|\psi\rangle = |\phi\rangle \implies \langle \psi | U^\dagger = \langle \phi | \; \; .
$$

\begin{definition}  \label{hermitian operators}
Hermitian operators are the ones that satisfy $U=U^\dagger$.
\end{definition}

A crucial set of unitary and hermitian operators in quantum mechanics are the Pauli matrices:
$$
X=\left(\begin{array}{ll}
0 & 1 \\
1 & 0
\end{array}\right) \quad Y=\left(\begin{array}{cc}
0 & -i \\
i & 0
\end{array}\right) \quad Z=\left(\begin{array}{cc}
1 & 0 \\
0 & -1
\end{array}\right) \; \; .
$$

They are used to represent the spin operators for spin-1/2 particles, the ones used for qubits as they have two possible states: spin up or down. Together with the identity matrix, the Pauli matrices form a very convenient basis for the space of 2x2 Hermitian matrices.

They possess numerous mathematical properties, such as the group they constitute, known as \textit{the Pauli group}. However, for this project, the crucial attribute we need is their ability to encode quantum operations on qubits.

Notice that qubit states $|0\rangle$ and $|1\rangle$ are eigenvectors of the Pauli-Z matrix, as $Z |0\rangle = |0\rangle$ and $Z |1\rangle = -|1\rangle$. That is why the basis $\{ |0\rangle, |1\rangle \}$ is also known as \textit{the Z-basis}, which will be extremely useful to encode binary variables in a Hamiltonian, as we will see in Section \ref{sec: QUBO intro}.

In the quantum computing gate model, this collection of matrices is crucial as their combinations enable the creation of many more quantum gates. These gates can then be implemented in a quantum circuit that manipulates an initial state to produce a desired output. However, we will not explore this paradigm of quantum computing; instead, we will focus on a different approach known as \textit{Adiabatic Quantum Computing} (AQC).

\subsubsection{Projection operators}
Another special family of operators that we will need is the \textit{projection operators}. Given an orthonormal basis $\{|\phi_i \rangle \}_{i=1, \ldots, N}$ we can define the operators $|\phi_i \rangle \langle \phi_i |$ which have the property that applied to any state $|\psi \rangle = \alpha_0 |\phi_0 \rangle + \ldots + \alpha_N |\phi_N \rangle$ we get:

\begin{align*}
|\phi_i \rangle \langle \phi_i |\psi \rangle &= 
|\phi_i \rangle \langle \phi_i |\left(\alpha_0 |\phi_0 \rangle + \ldots + \alpha_N |\phi_N \rangle\right) \\ &=
\alpha_0 |\phi_i \rangle \langle \phi_i |\phi_0 \rangle + \ldots +  \alpha_N |\phi_i \rangle \langle \phi_i |\phi_N \rangle \\
&= \alpha_i |\phi_i \rangle \; \; \text{(orthonormality of the basis)}.
\end{align*}

In other words, it projects the state vector $|\psi \rangle$ onto the direction given by the basis vector $|\phi_i \rangle$, giving us the coefficient that determines the probability of obtaining $|\phi_i \rangle$ when measuring $|\psi \rangle$ (the probability is $| \alpha_i |^2$). In general, a projection operator is any operator $A$ with the property $A^2=AA=A$.

Note that the result of this operator is not a valid quantum state, as the sum of the absolute squared values of its coefficients is given by $| \alpha_i |^2 = 1 -| \alpha_0 |^2-| \alpha_1 |^2- \ldots -| \alpha_{i-1} |^2-| \alpha_{i+1} |^2 - \ldots -| \alpha_N |^2$ which could differ from 1 if there exists any $\alpha_j \neq 0$, $j\neq i$. Consequently, these operators are not unitary, breaking the quantum properties of our system. However, they are essential for measuring a quantum state and obtaining a result at the end of any quantum algorithm.

\subsubsection{Hamiltonians}
\label{sec: hamiltonians}

We can give a more refined version of the second postulate of quantum mechanics which describes the evolution of a quantum system in continuous time \cite{nielsen00}.

\begin{post}
The time evolution of the state of a closed quantum system is described by the Schr{\"o}dinger equation,
$$
i \hbar \frac{d|\psi\rangle}{d t}=H|\psi\rangle \; \; .
$$
In this equation, $\hbar$ is a physical constant known as Planck's constant whose value must be experimentally determined. The exact value is not important to us. In practice, it is common to absorb the factor $\hbar$ into $H$, effectively setting $\hbar=1$. $H$ is a fixed Hermitian operator known as the Hamiltonian of the closed system.
\end{post}
For our purpose, we are interested in the time-independent Schr{\"o}dinger equation \cite{combarro2023practical}, which can be obtained by solving the differential equation using the separation of variables:
\begin{equation} \label{eq:time-indep eq}
    H|\psi\rangle = E |\psi\rangle
\end{equation}
where the Hamiltonian $H$ is a Hermitian matrix (an operator) that describes the total energy system, which does not depend on time. $|\psi\rangle$ is the wave function vector, which in the system described by $H$ has energy $E$ (a real value). Therefore, Equation \ref{eq:time-indep eq} is just an eigenvalue problem, and the fact that the matrix $H$ is Hermitian gives us useful information: 

\begin{itemize}
    \item $H$ is diagonalizable with real eigenvalues. That is why this matrix is hermitian by nature, because the energy must be a real number.
    \item It has $n$ linearly independent eigenvectors. It is always possible to find an orthonormal basis consisting of eigenvectors.
\end{itemize}

One final remark, in quantum physics the energy is quantized, i.e. there's a discrete set of possible energies in which the state can be when measured, and these are the eigenvalues of $H$. Moreover, we can obtain the \textit{expected energy} of a state that is not an eigenvector through the $H$ operator. Let $\{|\psi_1 \rangle, \ldots ,|\psi_N \rangle\}$ be a basis of the Hilbert space formed by eigenvectors of $H$, and let $|\phi \rangle = \alpha_1|\psi_1 \rangle + \ldots + \alpha_N|\psi_N \rangle$ be a non-elemental state (it has more than one coefficient different than $0$). Considering Equation \ref{eq:time-indep eq} we get
\begin{align*}
    H|\phi \rangle &= H\left(\alpha_1|\psi_1 \rangle+ \ldots + \alpha_N|\psi_N \rangle \right) \\
    &= \alpha_1H|\psi_1 \rangle+ \ldots + \alpha_NH|\psi_N \rangle \\
    &= \alpha_1E_1|\psi_1 \rangle+ \ldots + \alpha_NE_N|\psi_N \rangle && \text{(Equation \ref{eq:time-indep eq})} \; \; .
\end{align*}
\noindent If we now multiply by the bra vector, we end up with
\begin{align*}
    \langle\phi|H|\phi \rangle &= \left(\sum_i^n \alpha_i^* \langle \psi_i | \right) \left( \sum_i^n \alpha_i E_i | \psi_i \rangle \right)\\
    &= \alpha_1^*\alpha_1E_1\langle\psi_1|\psi_1 \rangle+ \ldots + \alpha_N^*\alpha_NE_N\langle\psi_N|\psi_N  \rangle && \text{(orthogonality)} \\
    &= |\alpha_1|^2 E_1 + \ldots + |\alpha_N|^2 E_N && \text{(orthonormality)}
\end{align*}
\noindent which is the probability of measuring each elemental state followed by its energy. Therefore, $\langle\phi|H|\phi \rangle$ is the \textit{expected energy} of state $|\phi \rangle$ after a measurement.

\section{QUBO formulation}
\label{sec: QUBO intro}

In this section, we will explore a highly convenient problem formulation for quantum computing, as it can be easily translated into the Hamiltonian of a quantum system. This conversion will be explained at the end of the section.

The Quadratic Unconstrained Binary Optimization (QUBO) is a combinatorial optimization problem that can be formulated as
$$
\min_{x}   x^T Q x
$$
where $x=(x_1, \ldots, x_N)^T \in \{0,1\}^N$ is a vector containing the $N$ binary variables of the problem and $Q=\{q_{ij}\}_{i,j\in\{1, \ldots, N\}}$ is a constant square matrix that depends on the problem formulation. The expression $x^T Q x$ is commonly known as \textit{cost function}, as it is a function of $x$ that could represent a cost we want to minimize. Components of $x$ are usually referred to as \textit{decision variables}, as they represent what ``decisions'' should be made to reach the global minimum, such as ``should item $i$ appear in the solution'' or ``should we take the road $j$ to reach the target city''.

As variables are binary we have the property that $x_ix_i = x_i$, and we can also redefine for $j \le i$ that $q_{ji}' := q_{ij} + q_{ji}$ and $q_{ij}' := 0$ , to save memory by making $Q$ triangular
$$
\min_{x} \sum^{N-1}_{i=1}\sum^N_{j>i} q_{ij}x_ix_j + \sum^{N}_{i=1} q_{ii}x_i \; \; .
$$

The formulation is unconstrained by definition, but we can encode equality constraints via penalty terms. Given the equality constraint
$$
\sum^{N}_{i=1} c_ix_i = C, \; c_i \in \mathbb{R}
$$
we can introduce it in the expression to minimize as
$$
\lambda \left( \sum^{N}_{i=1} c_ix_i - C\right)^2 \; \; ,
$$
where $\lambda \in \mathbb{R}$, known as \textit{Lagrange multiplier}, is a value that needs to be crafted to ensure that the constraint is always satisfied

On the other hand, to encode inequality constraints we would need to introduce new variables known as \textit{slack variables}. We can also introduce non-quadratic terms using these auxiliary variables. However, slack variables can considerably increase the dimension of the problem, which is a huge limitation in the current era of quantum computing where devices have a reduced number of qubits, so we will avoid this technique. In Section \ref{sec: ineq constraints} we will see a different approach, known as \textit{unbalanced penalization}, that allows us to introduce inequality constraints without increasing the number of variables.

Despite its simplicity, the QUBO formulation is incredibly powerful as it can encode a wide range of optimization problems, present in many fields like machine learning, finance, and logistics. What is more important for us though, is that it is extremely convenient for quantum computing because it can be easily translated into a Hamiltonian, so that the energy of our quantum system represents the value that we are trying to minimize, and the quantum state with that energy encodes the solution we are seeking. We call this quantum state of minimal energy the \textit{ground state}.

For this conversion, we just need the Pauli-Z matrix. In Section \ref{sec: unitary and hermitian operators} we saw that it satisfies $Z |0\rangle = |0\rangle$ and $Z |1\rangle = -|1\rangle$. As $Z$ is a Hermitian matrix, it could represent a Hamiltonian describing the energy of a system with just one spin 1/2 particle. If we now take a look into Equation \ref{eq:time-indep eq}, we see that such a particle would have a state $|0\rangle$ with energy 1 and a state $|1\rangle$ with energy -1. Therefore, this physical particle could represent what we know as \textit{spin variable}, a variable that can take two possible values: 1 and -1 (spin up or down).

We can trivially go from binary variables $x$ to spin variables $s$ by doing the transformation: 
$$
x = \frac{1-s}{2}
$$
so that $x=0 \iff s=1$ and $x=1 \iff s=-1$, which perfectly matches the mapping from states to energies. Now, given a cost function of a QUBO problem $f(\vec{x})$ we can obtain the expression of an equivalent problem but formulated with spin variables as $f(\frac{1-\vec{s}}{2})$ (change of variables performed element-wise on the vector).

Finally, we translate the spin cost function into a Hamiltonian operator by substituting spin variables $s_i$ for Pauli-Z matrices $Z_i$, where the subindex indicates that the Pauli-Z matrix is operating on the $i$-th qubit. In order to apply the operators into the correct elements of the basis explained in Section \ref{sec: hilbert space}, we need to use again the tensor product as follows
\begin{align*}
    Z_i &:= I \otimes \ldots \otimes Z_i \otimes \ldots \otimes I \\
    Z_i Z_j &:= I \otimes \ldots \otimes Z_i \otimes \ldots \otimes Z_j \otimes \ldots \otimes I
\end{align*}
where these products have $N$ terms, and $Z_i$ and $Z_j$ are placed in the $i$-th and $j$-th positions, respectively. We can rewrite the Hamiltonian obtained as
\begin{equation} \label{eq: Ising ham}
    H = -\sum_{i,j} J_{ij}Z_iZ_j -\sum_{j} h_i Z_i, \; \; J_{ij}, h_i \in \mathbb{R}
\end{equation}
which is the Hamiltonian of an Ising Model \cite{combarro2023practical}, that we will refer to when talking about quantum annealing in Section \ref{sec: QA}. In physics, we say that the first term contains the \textit{two-site interactions}, and the second term represents the \textit{external field}, which is trying to align each spin in the direction that $\vec{h}$ determines.

In conclusion, a quantum system with $N$ spin 1/2 particles described by a Hamiltonian derived from our QUBO formulation will have the property that states $|x_1 \ldots x_N \rangle$ (in the usual basis) have the energies given by our cost function $f(x_1, \ldots, x_N)$. Therefore, the QUBO problem becomes a problem of finding the quantum state with minimal energy, which allows us to unlock all the power of quantum computing to solve it.

\section{Adiabatic quantum computing}
\label{sec: AQC intro}

\textit{Adiabatic Quantum Computation} (AQC) is a quantum computing paradigm different from the gate model. It is based on the Adiabatic Theorem, which implies that if we consider a Hamiltonian that varies with time $H(t)$, whose time evolution will be described by the Schr{\"o}dinger equation, and we initially prepare the system in an eigenstate (such as the ground state), then the system will remain in the eigenstate during the time evolution as long as it is done ``sufficiently slow''.

This sets the idea of AQC. We consider the following time-dependent Hamiltonian
\begin{equation} \label{eq: AQC interpolation}
    H(t) = (1-s(t))H_0 + s(t)H_p \; \; \; \; \; \;  t\in [0, T], \; T\in \mathbb{R}_{>0}
\end{equation}
\noindent where $T$ is the duration of the evolution, $s(t)\in [0,1]$ is an increasing function known as the scheduling time that satisfies $s(0)=0$ and $s(T)=1$, $H_0$ is an initial Hamiltonian whose ground state is easy to prepare, and $H_p$ (known as the \textit{problem Hamiltonian}) is a Hamiltonian whose ground state encodes our solution. The goal is then to let the system evolve for as long as necessary so that at time $T$ it will be in the ground state of $H(1)=H_p$.

However, we have said that we will remain in the ground state during the evolution as long as it is done ``sufficiently slow''. To understand what is happening, remember that energy is quantized. If the evolution is too fast, there is a risk of the system leaping to a higher energy level. Therefore, the closer the second lowest energy (known as the \textit{first excited state}'s energy) is to the ground state's energy, the easier this energy jump is, so the slower the evolution needs to be to avoid it \cite{combarro2023practical}.

\begin{definition}  \label{min gap} 
The minimum gap is the minimum distance between the two lowest energies of the Hamiltonian at any time.
$$
g_{min} = \min_{s\in[0,1]} (E_1(s) - E_0(s))
$$
where $E_0(s)$ and $E_1(s)$ are the two lowest energy values of $H(s)$.
\end{definition}

If we define $s(t)=t/T, \; t\in [0,T]$, where $T\in \mathbb{R}_{>0}$ controls the interpolation rate \cite{Albash_2018}, then the evolution time needs to be greater than 
$$
T > \frac{1}{(g_{min})^2} \; \; .
$$
The issue here is that if the required time is too large (i.e. the minimum gap is too small), we could lose the properties of our quantum system before reaching the target Hamiltonian. This phenomenon is known as \textit{decoherence}, and occurs because quantum systems are extremely sensitive to environmental perturbations. That's why the study of the minimum gap is crucial to make AQC work \cite{combarro2023practical}.

Finally, AQC is polynomially equivalent to other quantum computing models \cite{combarro2023practical}, such as the gate-based model. This means that any problem that can be solved with a regular quantum computer can also be tackled using AQC.

\subsection{Quantum annealing}
\label{sec: QA}

\textit{Quantum annealing (QA)} is an algorithm that relies on the same idea as AQC to solve optimization problems encoded as the ground state of a Hamiltonian. It is implemented in a special type of quantum computer called \textit{quantum annealer} \cite{combarro2023practical}. 

The optimization starts in an equal superposition of all possible states, i.e. all possible solution candidates. We achieve this by considering in Equation \ref{eq: AQC interpolation} the following initial Hamiltonian
$$
H_0 = - \sum_{i=1}^N X_i
$$
where $X_i$ is the Pauli-X matrix which has an eigenvector $(1,1)^T$ of eigenvalue $1$ and an eigenvector $(-1,1)^T$ of eigenvalue $-1$. Therefore, the ground state of $H_0$ (after normalization to have a valid quantum state) would be $| \psi \rangle = \frac{1}{\sqrt{2}}| 0 \rangle + \frac{1}{\sqrt{2}}| 1 \rangle$. As $H_0$ applies this operator to each qubit, the resulting ground state is an equal superposition of all possible states, i.e. if measured it could collapse to any element of the canonical basis with equal probability.

Once we have the initial Hamiltonian with an easy-to-prepare ground state, we need to define the problem Hamiltonian that encodes the solution we seek. Typically, due to both hardware limitations and the versatility of the model, in quantum annealers the problem Hamiltonian $H_p$ is assumed to be an Ising model Hamiltonian (Equation \ref{eq: Ising ham}), which are the ones we obtained from the QUBO formulation in Section \ref{sec: QUBO intro}. This class of Hamiltonians is not believed to be sufficient to build a universal adiabatic quantum computer (in particular, $H(t)$ belongs to what is called stoquastic Hamiltonians \cite{Albash_2018}), but we can solve a wide range of problems with it.

Apart from this limitation, in quantum annealing the evolution is not guaranteed to be done sufficiently slow. This is because accurately calculating the minimum gap required to compute the necessary evolution time is difficult, and even if we could, the required time could be impractically long. QA can still yield good solutions, but remaining in the ground state (i.e. the optimal solution) until the end of the evolution is not guaranteed \cite{combarro2023practical}.

Due to these limitations, QA is not universal like AQC. Some problems solvable by other quantum computers cannot be tackled by quantum annealers. However, quantum annealers are still very useful, especially because they are relatively easy to build with a large number of qubits \cite{combarro2023practical}.

%% file: chapters/C2_TN.tex
\chapter{Introduction to tensor networks}

Tensor networks are a powerful mathematical tool to represent and manipulate complex systems efficiently \cite{Bridgeman_2017, TNorg, DMRG}. They have found applications across diverse fields, from quantum information theory \cite{Or_s_2014} to machine learning and beyond \cite{GlenEvenbly2022practical, TNBasicIntro}. In this introduction, we will explore the fundamental concepts required to represent and manipulate quantum many-body systems with them, a topic we'll delve into in the next chapter.

In this chapter, we will introduce the fundamental concepts of tensor Networks, starting with the main definitions and the diagrammatic notation, and covering important topics like some of the available operations we can perform with them. Finally, we will explain the Singular Value Decomposition (SVD), and discuss its significance in this context. 

\section{Basic concepts: notation, operations, and gauge freedom}

A nice way to get introduced to tensors is to think about them as a generalization of a vector and a matrix. A vector is just a 1-rank tensor, as it has 1 index, a matrix is a 2-rank tensor, as we need two indices to cover it, but nothing stops us from adding more indices, in order to get higher dimension tensors.

\begin{definition}  \label{tensor_def}  

A tensor is an algebraic object that can be represented as a multidimensional array that obeys certain transformation rules. An $n$-th rank tensor is a tensor that has $n$ indices $i_1,...,i_n$, each one with dimensions $d_1,..., d_n$. It would then have $d_1 \cdot ... \cdot d_n$ components

\end{definition}
The notation is similar to the one we use with matrices. We can use upper and lower indices indistinctively, but, we will make a distinction between them in the next section, as \textit{physical} and \textit{virtual} indices.

\begin{figure}[ht]
  \centering
  \subfloat[Tensor examples.]{\includegraphics[width=0.4\textwidth]{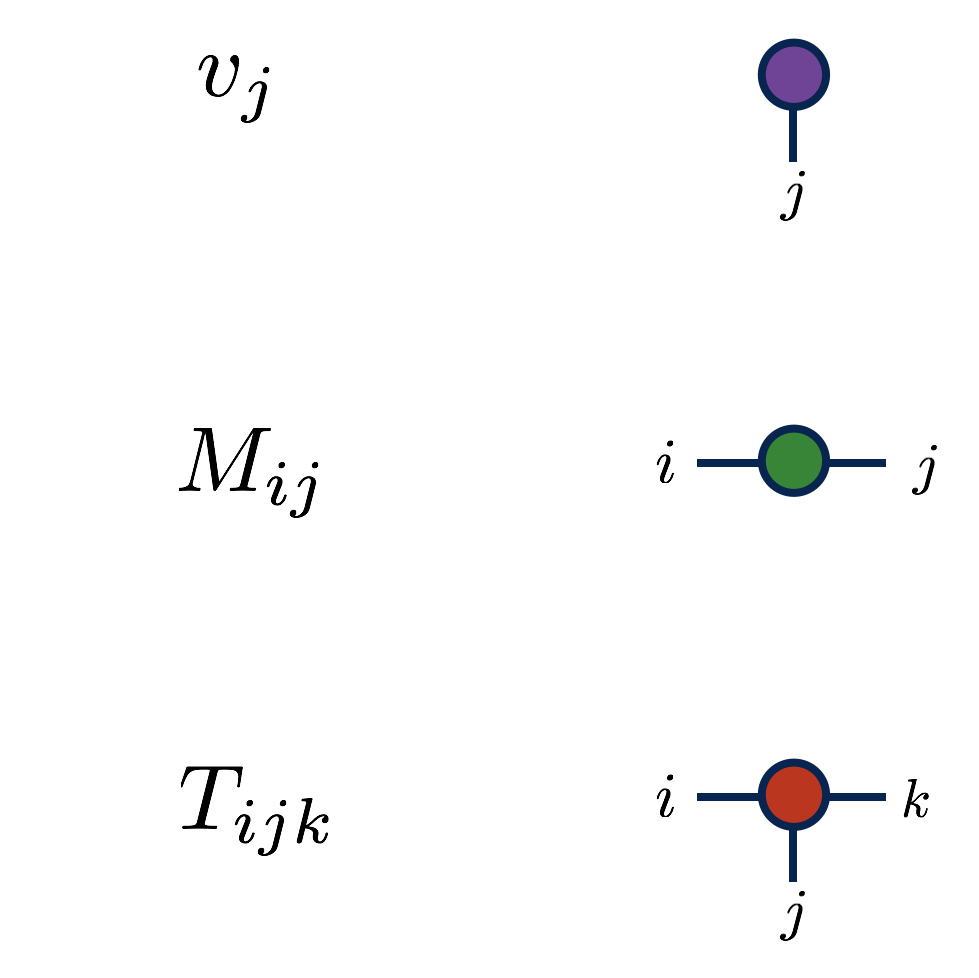}\label{fig: diagram f1}}
  \hfill
  \subfloat[Identity matrix.]{\includegraphics[width=0.3\textwidth]{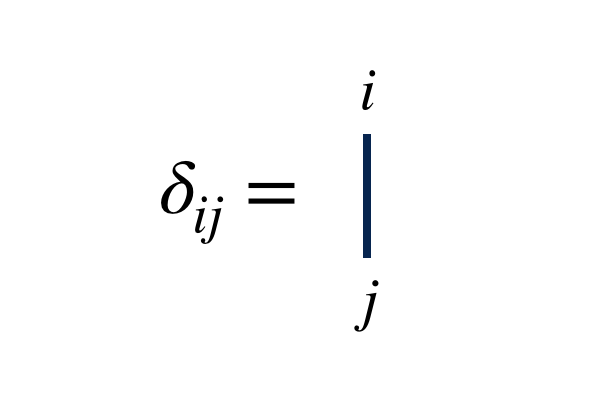}\label{fig: diagram f2}}
  \caption{(a) shows the diagrammatic representation of a vector (1-rank tensor), a matrix (2-rank tensor), and a 3-rank tensor, while (b) illustrates the notation for the identity matrix in tensor notation. Images are from Ref. \cite{TNorg}.}
  \label{fig: basic tensor examples}
\end{figure}

The most convenient way to represent a tensor is through diagrams. We use shapes (usually filled) to represent tensors, and lines emanating from these shapes to represent its indices. The representation of a vector, a matrix, and a 3-rank tensor in diagrammatic notation is the one illustrated in Fig. \ref{fig: diagram f1}.

\subsubsection{Special types of tensors}

We will use a special notation to design two special types of tensors: identity and diagonal matrices. Identity matrices (described by the Kronecker delta) in tensor notation are just a plain line, as depicted in Fig. \ref{fig: diagram f2}. Adding a dot to this line represents a diagonal matrix (i.e. $\lambda_i \delta_{ij}$).

Another special tensor that we will need is the unitary tensor, which extends the notion of a unitary matrix.

\begin{definition}  \label{unitary tensor}
A tensor $U$ is unitary if there exists a bipartition of the indices under which the tensor could be reshaped into a unitary matrix 
\end{definition}

\subsubsection{Tensor operations}
As we said before, tensors obey certain transformation rules. There are essentially 3 main operations we can perform: \textit{permutation}, \textit{reshaping}, and \textit{contraction}.

A permutation of a tensor involves merely rearranging the order of its indices. The tensor remains with the same components; it is just a matter of changing the labels assigned to each index. On the other hand, reshaping combines one or more indices into a single larger one, or vice-versa. Consequently, reshaping alters the number of indices but not the total dimension.

Lastly, we have contractions, which again can be viewed as a generalization of the matrix product. In a matrix product, we sum over some specific indices (the column index of the left matrix with the row index of the right matrix), but nothing stops us from summing over any other index if the dimensions fit.

\begin{figure}[H]
    \centering
    \includegraphics[width= 0.8\columnwidth]{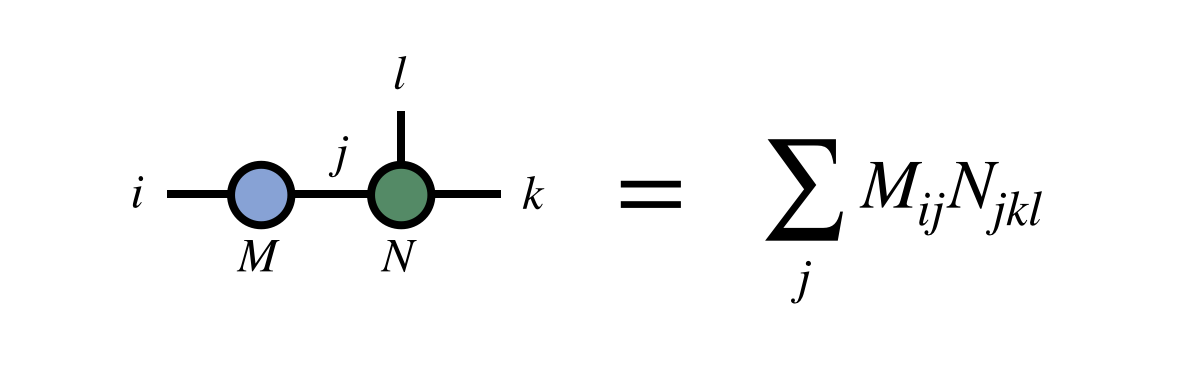}
    \caption{Example of the contraction of a 2-rank tensor with a 3-rank tensor over a common index. Image from Ref. \cite{TNorg}.}
    \label{fig: contraction}
\end{figure}

When we connect multiple tensors via contractions we obtain a tensor network. The best way to work with them is via diagrams, where connecting two index lines implies a contraction or summation over the connected indices. Let's say we have a 2-rank tensor $M$ with indices $i,j$, and a 3-rank tensor $N$ with indices $j,l,k$. Performing a contraction over the common index $j$ would be the operation shown in Fig. \ref{fig: contraction}.

We sum over the $j$ index and the result would be a 3-rank tensor with indices $i,l,k$. It does not matter how complicated the tensor network is, the resulting tensor after performing all the contractions will have a rank equal to the number of open indices (i.e. unpaired lines).

In general, the computational cost of contractions depends on the dimensions of both open and contracted indices. Let $j_1, \ldots j_n$ be the contracted indices (just counted once), $k_1, \ldots, k_m$ be the open indices, and $d_i$ be the dimension of index $i$. Then, the contraction would have a cost
$$
O\left(\prod_{j=1}^n d_{j_i} \cdot \prod_{i=1}^m d_{k_i}\right)
$$

For example, if we look at Fig. \ref{fig: contraction} and we suppose that every index has dimension $d$, then the cost would be $O(d^4)$. However, there could be more efficient implementations for some special cases. For example, if we take the previous formula then the cost of matrix multiplication of two $d\times d$ matrices (which is a particular case of tensor contraction) would be $O(d^3)$, but there exists more efficient implementations that take $O(d^{2.373})$ \cite{TNBasicIntro}.

\begin{figure}[H]
    \centering
    \includegraphics[width= 0.8\columnwidth]{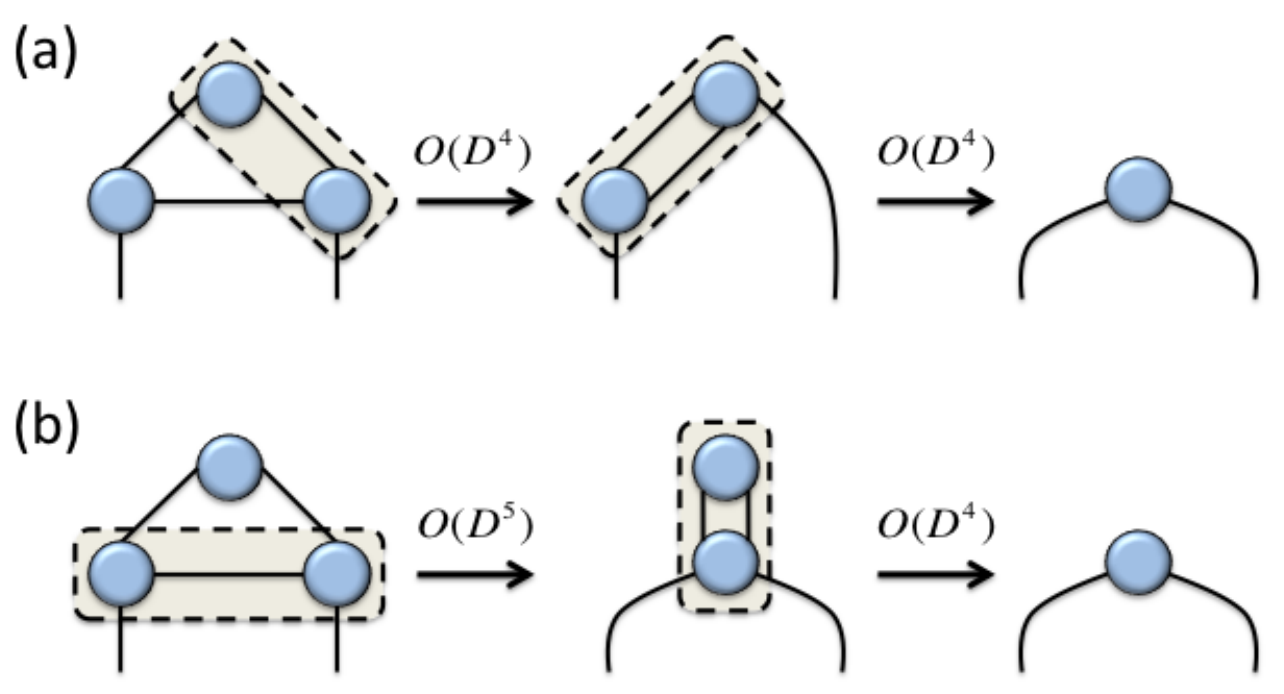}
    \caption{Image from \cite{Or_s_2014} showing the importance of the order in which indices are contracted in the contraction of multiple tensors, assuming each index has a dimension $D$. See how case (a) is more efficient because it starts the contraction with less interconnected tensors, but this heuristic is not always right (in fact, determining the optimal ordering is an NP-hard problem \cite{GlenEvenbly2022practical}).}
    \label{fig: order matters}
\end{figure}

Finally, notice that this computational cost implies that the order in which we contract indices when there are more than two tensors involved matters. The diagram in Fig. \ref{fig: order matters} greatly exemplifies this fact.

\subsubsection{Gauge Freedom}
\label{sec: gauge freedom}

Let $T$ be a tensor network that, under the contraction of all internal indices, evaluates to some tensor $D$. We can then see $T$ as a decomposition of $D$ in two or more tensors. A natural question should arise to any mathematician: is this decomposition unique? Is there a different TN $T'$ with the same network geometry that also evaluates to $D$? The answer is clearly yes, as we can introduce any invertible tensor $X$ and its inverse $X^{-1}$ on any edge of the TN and absorb them into the neighboring tensors, leaving the final evaluation $D$ unchanged.

We call this lack of uniqueness the \textit{Gauge Freedom}. However, we can still look for canonical forms that offer advantages for performing certain calculations. Tensor network algorithms exploit these forms to do more efficient computations. A canonical form we will employ in the next chapter is the notion of \textit{orthogonality} \cite{GlenEvenbly2022practical}.

\begin{definition} \label{def: orthogonal}
A tensor in a tensor network is considered an orthogonality center if each edge connected to it simplifies to the identity when contracted with its conjugate.
\end{definition}

We will use this definition in the next chapter.

\section{The Singular Value Decomposition}
\label{sec: SVD}

We have observed how contraction enables the merging of tensors. However, it is equally interesting to split tensors into simpler blocks for more efficient computations. This is precisely the role of the \textit{Singular Value Decomposition} (SVD): a crucial tool for compressing and optimizing tensor networks.

SVD is a matrix decomposition technique, but any rank-N tensor can be interpreted as a matrix through the reshape operation, transforming it into a rank-2 tensor. According to SVD \cite{DMRG}, any matrix $M \in \mathbb{C}^{m \times n}$ can be decomposed as

\begin{equation} \label{eq: SVD}
    M=U S V^{\dagger}
\end{equation}

where
\begin{itemize}
    \item $U \in \mathbb{C}^{m \times \min (m, n)}$ has orthonormal column vectors (i.e. they have norm one and are mutually orthogonal with the standard inner product in the vector space given)
    \item $S = diag(s_1, s_2, ..., s_r,0,...,0) \in \mathbb{R}_{\ge 0}^{\min (m, n) \times \min (m, n)}$, $r\le min(m,n)$ where $s_1 \ge s_2 \ge  ... \ge  s_r$ are called \textit{singular values of M} (and $r$ is the rank of $M$).
    \item $V^{\dagger} \in \mathbb{C}^{\min (m, n) \times n}$ has orthonormal row vectors.
\end{itemize}

Would $U$ (or $V^{ \dagger}$) also be square matrices, they would also be unitary, as $U U^{\dagger}=I$ ($V V^{\dagger}=I$). Also note that the $U$ matrix in the decomposition is in the left-orthogonal form, and $V^{\dagger}$ is in the right-orthogonal form. SVD can also be defined with $U \in \mathbb{C}^{m \times m}$, $S \in \mathbb{R}_{\ge 0}^{m \times n}$, and $V \in \mathbb{C}^{n \times n}$. This formulation could be less efficient in certain situations as it could contain more elements, but it will not be a concern for our purpose.

But why do we care about SVD? Because it provides the solution to the \textit{low-rank matrix approximation problem}: given a matrix $M$, we seek a rank-$k$ matrix $M_k$, $k<$ rank $M$, that minimizes $|| M-M_k ||_F$, i.e. $M_k$ has the property that
$$
|| M-M_k ||_F \le || M-A ||_F \; \; \; \forall A \mid \text{rank}(A) = k
$$
where $F$ is the Frobenius norm, which is defined for a matrix $A$ of size $m\times n$ as
$$
|| A ||_F = \sqrt{\sum^m_i\sum^n_j |a_{i,j}|^2} \; \; \; .
$$

Turns out that the rows/columns corresponding to smaller singular values of $M$ have a lesser impact on the Frobenius norm of the matrix \cite{Bridgeman_2017}. Therefore, the best approximation $M_k$ is achieved by setting the smallest singular values in the SVD of $M$ (i.e. in the $S$ matrix) to zero, keeping only the $k$ most significant ones. We end up with

\begin{align*} 
   & M_k=U S_k V^{\dagger} && \text{where $S_k = diag(s_1, s_2, ..., s_k,0,...,0)$, $k<$ rank $M$} \; \; \; .
\end{align*}

This technique is called \textit{truncation} and is widely used to simplify the complexity of TN. We will see the implications of this in the next chapter.

%% file: chapters/C3_Connection_TN_Quantum.tex
\newcounter{chapterthreefootnote}
\setcounter{chapterthreefootnote}{1} 

\newcommand{\chapterthreefootnotemark}{\footnotemark[\value{chapterthreefootnote}]}
\newcommand{\chapterthreefootnotetext}{\footnotetext[\value{chapterthreefootnote}]{All images in Chapter 3 are from Ref. \cite{TNorg}\label{chap3footnote}}}

\chapter{Connection between tensor networks and quantum mechanics}

One of the challenges in quantum simulation is that the Hilbert space is exponentially large, specifically $d^N$, where $N$ is the number of particles and $d$ is the number of possible states for each particle \cite{Bridgeman_2017}. For example, simulating a caffeine molecule would require $10^{48}$ classical bits to be simulated, which is infeasible considering there are $10^{50}$ atoms in the Earth.

Performing exact diagonalization on the Hamiltonian matrix would tell us almost everything about our quantum system. However, the cost of diagonalization scales exponentially with the matrix size \cite{TNBasicIntro}. An alternative approach is using \textit{Monte Carlo} methods, which can yield accurate solutions in polynomial time. Unfortunately, certain scenarios suffer from the so-called \textit{sign problem}, making these methods incredibly challenging to apply for certain systems \cite{TNBasicIntro}.

Quantum computers, on the other hand, are well-suited for managing simulations and other computations in the Hilbert space because they leverage quantum mechanics for calculations \cite{nielsen00}. However, these are currently theoretical results, as quantum computers are still in an early stage and lack the computing power to solve practical problems. That is where TN come into play \cite{Bridgeman_2017}.

In this chapter, we will explore how TN are used to describe quantum systems \cite{video_lecture_DMRG, cylinder, AppendixA, automataIntro, Or_s_2014, TNBasicIntro, DMRG}. This mathematical framework naturally represents quantum many-body systems with a number of parameters that scale polynomially instead of exponentially. After that, we will use this convenient representation to run the \textit{Density matrix renormalization group} (DMRG) algorithm to approximate the lowest energy state of a quantum system. Finally, we will introduce a modification on this algorithm to compute the first excited state, in order to approximate the minimum gap of an annealing Hamiltonian.

\section{A natural representation of quantum systems}
\label{sec: TN representation of quantum systems}

Let's consider a quantum many-body system of $N$ particles. Let the physical dimension of the $k$-th particle (or site) be $d_k$. This means that the degrees of freedom of each one of these particles can be described by $d_k$ different states, that we can denote $\{ \left| 0 \right>, \left| 1 \right>, ..., \left| d_k -1 \right> \}$ (for instance, d = 2 for a qubit or spin 1/2 particle). Therefore, the wave function that describes the physical properties of this system will be:
\begin{align*}
    |\psi\rangle&=\sum_{i_1 i_2 \ldots i_N} C_{i_1 i_2 \ldots i_N}\left|i_1 \ldots i_N\right\rangle\; , \; \; \; &&i_k = 0,..., d_k -1, \; \; \; k = 1, \ldots , N \; \; .
\end{align*}
This is just a sum of tensor products of local basis vectors. The coefficients $C_{i_1 i_2 \ldots i_N} \in \mathbb{C}$ only need to fulfill the constraint that $|\psi\rangle$ is normalized
$$\sum_{i_1 i_2 \ldots i_N} |C_{i_1 i_2 \ldots i_N}|^2 = 1 \; \; .$$

Therefore, these $\prod_i d_i$ complex numbers represent all the information of our quantum system, and here is where we can unleash the power of TN: we can interpret the coefficients $C_{i_1 i_2 \ldots i_N}$ as a tensor with $N$ indices, each of them corresponding to a particle $k$ (index with dimension $d_k$).

\begin{figure}[H]
    \centering
    \includegraphics[width= 0.75\columnwidth]{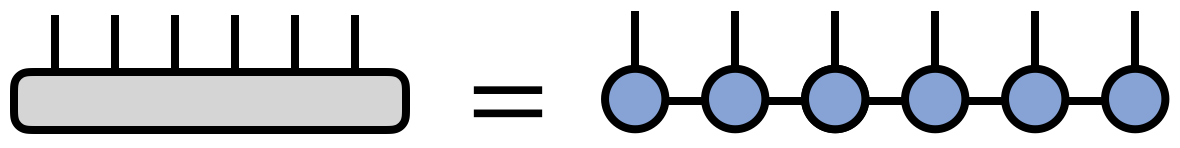}
    \caption{Decomposition of the initial tensor $C_{i_1 i_2 \ldots i_N}$ (left), which is an $N$-rank tensor, into its MPS form (right), which contains $N-2$ rank-3 tensors and two boundary rank-2 tensors.\protect\chapterthreefootnotemark}
    \label{fig: MPS}
\end{figure}

Now, this tensor representation has an issue: we said that the Hilbert space is huge, and so will be our tensor. However, instead of the tensor $C$ we can use its \textit{Matrix Product State} (MPS) form to obtain a more efficient representation
\begin{align} \label{Eq: MPS}
    |\psi\rangle&=\sum_{i_1 i_2 \ldots i_N} M_{i_1} \ldots M_{i_N} \left|i_1 \ldots i_N\right\rangle \; , \; \; \; &&i_k = 0,..., d_k -1, \; \; \; k = 1, \ldots , N \; \; ,
\end{align}
\noindent where $M_{i_k}$ are matrices with dimensions
$$(1\times m_1), (m_1\times m_2), \ldots ,(m_{N-2}\times m_{N-1}), (m_{N-1}\times 1) \; \; .$$
Dimensions $m_i$ are known as the \textit{bond dimension} between sites $i$ and $i+1$. This gives us a state representation with $O(\sum_i \m_{i-1} \cdot m_{i})$ many parameters. Remember that this representation is not unique, we can change it by introducing any invertible matrix $X$ s.t. $XX^{-1}=I$ and absorbing it into two consecutive matrices. A visual representation of an MPS can be found in Fig. \ref{fig: MPS}.
\chapterthreefootnotetext

Notice that the left representation in Fig. \ref{fig: MPS} contains $d_1d_2 \ldots d_N$ parameters, while the right representation has $d_1m_1 + m_1d_2m_2 + \ldots + m_{N-2}d_{N-2}m_{N-1} + m_{N-1}d_N$. Therefore, the number of parameters increases linearly with the number of particles instead of exponentially. The trade-off here is that, in general, that would require exponentially large bond dimensions, but we will see how to limit them in Section \ref{sec: MPS conversion}.

{\bf Notation:} In the diagram of an MPS (see Fig. \ref{fig: MPS}), the vertical lines are referred to as \textit{physical indices}, as they represent the index associated with each particle, while the horizontal lines are called \textit{virtual indices}.

\subsubsection{MPS orthogonality}
\label{sec: orthogonality}

Remember how we mentioned the notion of orthogonality center in Definition \ref{def: orthogonal}. In the case of MPS, it will be useful to make a further distinction.

\begin{figure}[H] 
    \centering
    \includegraphics[width= 0.4\columnwidth]{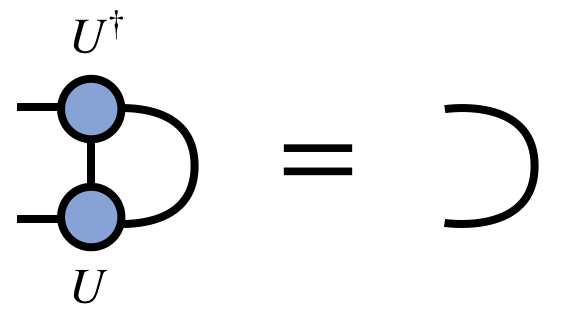}
    \caption{Right-orthogonal tensor in an MPS.\protect\chapterthreefootnotemark}
    \label{fig: ortho}
\end{figure}

\begin{definition}  \label{Tensor orthogonality}
Given a tensor $U$ from an MPS, i.e. it has one physical index and two virtual ones, we say that it is in a \textit{right-orthogonal form} if it satisfies the property shown in Fig. \ref{fig: ortho}.
\end{definition}
\chapterthreefootnotetext

{\bf Notation:} When physical indices are pointing down, like in Fig. \ref{fig: ortho}, it will represent the complex conjugate, as shown in the definition. Consequently, an MPS with all its physical indices pointing down will represent the bra vector of the state, allowing for a more concise representation of the inner product of two states by connecting the two MPS from their physical indices and contracting the entire tensor network.

If we reshape to join the two indices being contracted we obtain a matrix product $U U^\dagger = I$. In other words, the tensor, if contracted by the appropriate indices, is a unitary matrix.

Analogously, we say that the tensor is in a \textit{left-orthogonal form} if we obtain the same result doing the contraction over the indices at the left ($V^\dagger V = I$). We will see that we can obtain tensors with these properties using the SVD.

\begin{definition}  \label{MPS orthogonality}
We say that an MPS is in a right (resp. left) orthogonal form if all tensors in the network except for the first (resp. last) one are right (resp. left) orthogonal. Diagrammatically, an MPS in a right-orthogonal form is displayed in Fig. \ref{fig: isometric_U}.
\begin{figure}[H]
    \centering
    \includegraphics[width= 0.75\columnwidth]{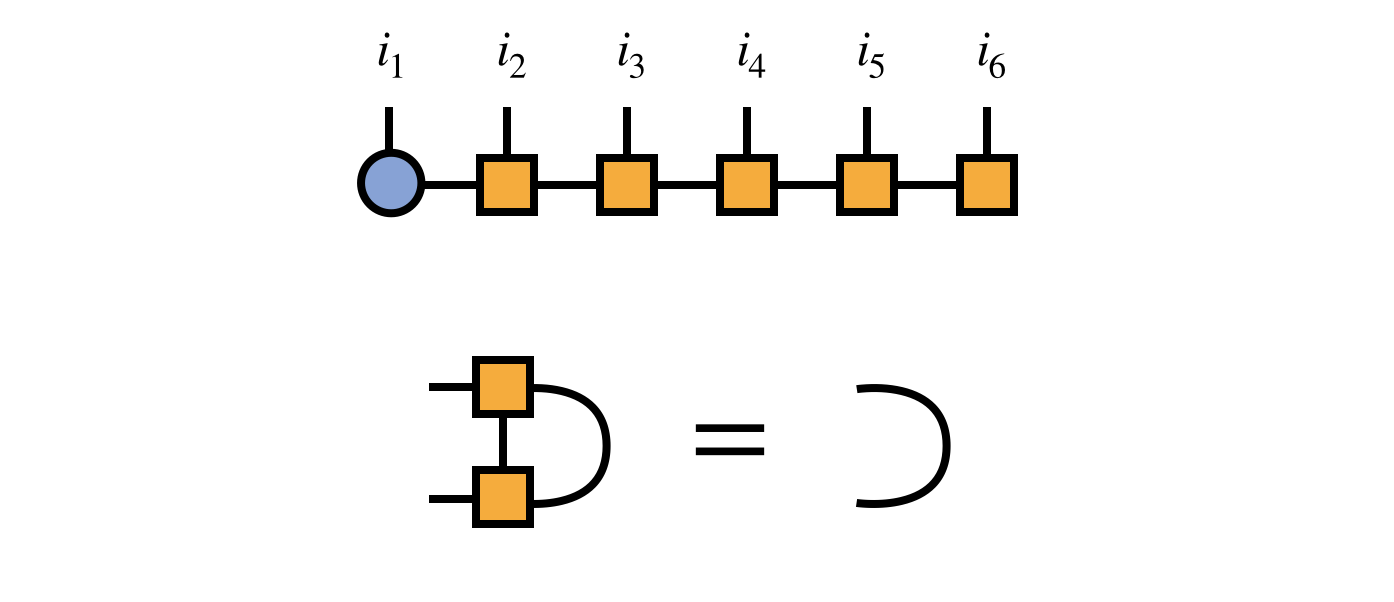}
    \caption{MPS in a right-orthogonal form.\protect\chapterthreefootnotemark}
    \label{fig: isometric_U}
\end{figure}
\end{definition}
\chapterthreefootnotetext

{\bf Notation:} Orthogonal tensors are usually represented by a square, as shown in Fig. \ref{fig: isometric_U}.

These properties are fundamental as they can greatly simplify the contraction of a TN, as we will see in Section \ref{sec: DMRG} when explaining the DMRG algorithm.

\subsubsection{Conversion from a general tensor to its MPS form}
\label{sec: MPS conversion}

The standard method for obtaining the MPS form of a rank-N tensor is through the SVD, which produces an MPS in the orthogonal form outlined before.

Let's consider a rank-$N$ tensor $C_{i_1, \ldots, i_N}$. By reshaping its indices, we can view it as a matrix $C_{i_1, (i_2, \ldots, i_N)}$ of dimensions $\left(d_1 \times (d_2\cdot \ldots \cdot d_N) \right)$. With this matrix at hand, we can proceed to perform its SVD
$$
C_{i_1, (i_2, \ldots, i_N)} = \sum_{a_1} U_{i_1, a_1} S_{a_1, a_1} V_{a_1, (i_2 \ldots i_N)}^{\dagger} \; \; .
$$
If we define
\begin{align*}
    & M_{i_1} := U_{i_1, a_1} && C_{a_1, (i_2, \ldots, i_N)}:= S_{a_1, a_1} V_{a_1, i_2 \ldots i_N}^{\dagger} \; \; ,
\end{align*}
we just obtained the first tensor of the MPS in Equation \ref{Eq: MPS}. Note that $a_1$ is the virtual index connecting $M_{i_1}$ with the remaining part of the original tensor to be decomposed. We have to take extra care with this index for the next iteration. We reshape $C_{(a_1,i_2),(i_3\ldots, i_N)}$ and do SVD again to obtain
$$
C_{(a_1,i_2),(i_3\ldots, i_N)} = \sum_{a_2} U_{(a_1,i_2), a_2} S_{a_2, a_2} V_{a_2, (i_3 \ldots i_N)}^{\dagger} 
$$
\begin{align*}
    & M_{i_2} := U_{a_1,i_2,a_2} && C_{a_2, (i_3, \ldots, i_N)}:= S_{a_2, a_2} V_{a_2, i_3 \ldots i_N}^{\dagger} \; \; .
\end{align*}

\begin{figure}[H]
    \centering
    \includegraphics[width= 0.8\columnwidth]{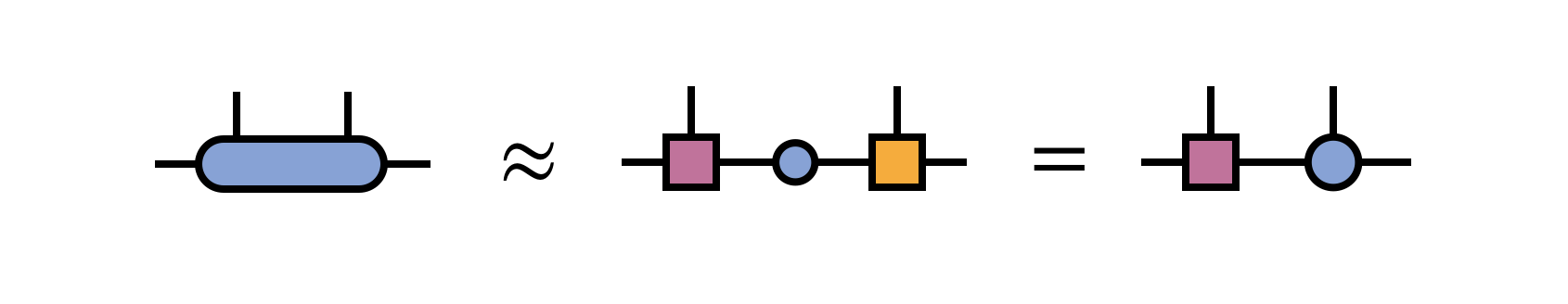}
    \caption{Truncation of a 4-rank tensor into its MPS form. Through the reshape operation it can be considered as a matrix to apply the SVD, and then with truncation we can control the size of the bond dimension. Note that when absorbing $S$ into $V$ we lose the orthonormality of its rows.\protect\chapterthreefootnotemark}
    \label{fig: SVD}
\end{figure}

Now $M_{i_2}$ has three indices: the previous virtual index $a_1$, the physical index $i_2$, and the new virtual index shared with the remaining part of the original tensor. We can also limit the bond dimension (i.e. the dimension of the virtual indices $a_i$) by employing the truncation technique explained in Section \ref{sec: SVD} after each SVD operation. Typically, we denote the upper bound set as $\chi$. See Fig. \ref{fig: SVD} to see the process diagrammatically.

\chapterthreefootnotetext

We can repeat this process to separate each physical index until we get
\begin{align*}
    C_{i_1,i_2, \ldots, i_N} & = M_{i_1} \ldots M_{i_N} \\
    & = \sum_{a_1 \ldots a_{N-1}} U_{i_1, a_1} U_{a_1,i_2,a_2} \ldots U_{a_{N-2},i_{N-1},a_{N-1}} C_{a_{N-1},i_N} \; \; \;,
\end{align*}
which is precisely the MPS form of Equation \ref{Eq: MPS}.

It's worth noting that the tensors we obtain (except for the last one) are the $U$ matrices from the SVD, which implies that they have orthonormal column vectors. Consequently, the resulting MPS is in a left-orthogonal form, which will be very advantageous for executing the DMRG algorithm. If a right-orthogonal form is needed instead, we simply need to reverse the process to create the MPS from the $V$ matrices of the SVD.

\subsubsection{The role of entanglement in MPS representation}

We have just seen that any state can be written as a tensor, and any tensor can be written in an MPS form, which means that any state can be represented as an MPS. Unfortunately, for some states, the bond dimension required can be extremely large, making the truncation result in a drastically different MPS. Therefore, the final representation's quality will depend on the number and size of the non-zero singular values we have to truncate.

To identify which states are less impacted by truncation, we have to look at the entanglement that the state presents \cite{DMRG}. Qualitatively, the lower the entanglement, the better the representation is because there is a strong relation between the entanglement of a state and the singular values we obtain when computing its MPS.

To justify this we need first to introduce a measurement of the entanglement of a system \cite{nielsen00}, and then see how it relates to the singular values obtained in the creation of the MPS.

\begin{theorem}{\bf (Schmidt decomposition)} \label{thm: Schmidt decomposition} 
Given a bipartition of a Hilbert space $\mathcal{H} = \mathcal{H}_A \otimes \mathcal{H}_B$, any state $|\psi\rangle \in \mathcal{H}$ can be written as a \textit{Schmidt decomposition} into orthonormal vectors $|\alpha_1 \rangle, \ldots , |\alpha_n \rangle \in \mathcal{H}_A$ and $|\beta_1 \rangle, \ldots |\beta_m \rangle \in \mathcal{H}_B$ as
$$
| \psi \rangle = \sum_i s_i |\alpha_i \rangle \otimes |\beta_i \rangle
$$
where $s_i$ are called \textit{Schmidt coefficients}. These coefficients can be chosen as real and non-negative and need to satisfy
$$
\langle \psi | \psi \rangle = \sum_i |s_i|^2 = 1  \; \; \; .
$$
\end{theorem}

Before proving this theorem let's define the entanglement of a state to see how all bounds together.

\begin{definition}  \label{Entanglement entropy} 
Given a bipartition of a Hilbert space $\mathcal{H} = \mathcal{H}_A \otimes \mathcal{H}_B$, and a state $| \psi \rangle\in \mathcal{H}$, the \textit{entanglement entropy} of one part with the other is given by
$$
S(A) = S(B) = -\sum_i |s_i|^2 \log |s_i|^2
$$
where $s_i$ are the Schmidt coefficients of $| \psi \rangle$.
\end{definition}

Therefore, entanglement is determined by the Schmidt coefficients: the fewer and smaller these Schmidt coefficients are, the lower the entanglement of the state is. Finally, the relation of Schmidt coefficients with singular values is displayed in the proof of Theorem \ref{thm: Schmidt decomposition}.

\begin{proof}[Proof of Theorem \ref{thm: Schmidt decomposition}]
Let $\{|a_1\rangle, \ldots, |a_n\rangle\}$ be an orthonormal basis of $\mathcal{H_A}$ and $\{|b_1\rangle, \ldots, |b_m\rangle\}$ be an orthonormal basis of $\mathcal{H_B}$. We can write $|\psi\rangle$ as 

$$
|\psi\rangle = \sum_{i=1}^n \sum_{j=1}^m C_{ij} |a_i\rangle \otimes |b_j\rangle
$$

Notice that $C=\left(C_{ij}\right)_{i,j}$ is a $n\times m$ matrix. Therefore, we can consider its SVD

\begin{align*}
    &C = U S V^{\dagger} && U \in \mathbb{C}^{n \times l}, \; S \in \mathbb{R}^{l \times l}, \; V^{\dagger} \in \mathbb{C}^{l \times m}
\end{align*}

where $l=\min(n,m)$. Let $\{|u_1\rangle, \ldots, |u_l\rangle\}$ be the orthonormal column vectors of $U$ and $\{|v_1\rangle, \ldots, |v_l\rangle\}$ be the column vectors of $V$, which implies that $\{\langle v_1|, \ldots, \langle v_l|\}$ are the orthonormal row vectors of $V^{\dagger}$. 

Then, we can rewrite the decomposition as
$$
C = \sum_{k=1}^l s_k |u_k\rangle \langle v_k |
$$
where $s_i$ are the singular values. Then, each element would be
$$
C_{ij} = \sum_{k=1}^l s_k |u_k\rangle_i \langle v_k|_j
$$

$$
|\psi\rangle = \sum_{i=1}^n \sum_{j=1}^m \sum_{k=1}^l s_k |u_k\rangle_i \langle v_k|_j |a_i\rangle \otimes |b_j\rangle
$$

Writing $| \alpha_k \rangle = \sum_{i=1}^n |u_k\rangle_i |a_i\rangle$ and $| \beta_k \rangle =\sum_{j=1}^m \langle v_k|_j |b_j\rangle$ we get

$$
|\psi\rangle = \sum_{k=1}^l s_k | \alpha_k \rangle \otimes | \beta_k \rangle
$$

Finally, from the fact that $U$ is unitary and the orthonormality of $|a_i\rangle$ it can be deduced that $| \alpha_k \rangle$ forms an orthonormal basis of $\mathcal{H_A}$ \cite{nielsen00}. Same applies for $| \beta_k \rangle$ as a basis of $\mathcal{H_B}$.

\end{proof}

This proof shows the relation we were looking for: Schmidt coefficients are equivalent to singular values. This means that if the entanglement entropy is smaller, the spectrum of Schmidt coefficients decays faster to zero \cite{DMRG}, and so does the singular values computed during the creation of the MPS. Consequently, the truncation (i.e., the deletion of the least significant singular values) needed for states with low entanglement will have a lower impact in the accuracy of the MPS representation, implying that $\chi$ can be substantially reduced without compromising the accuracy of the representation.

\subsubsection{A natural representation of operators}

\begin{figure}[ht]
  \centering
  \subfloat[Operator written as a tensor.]{\includegraphics[width= 0.5\columnwidth]{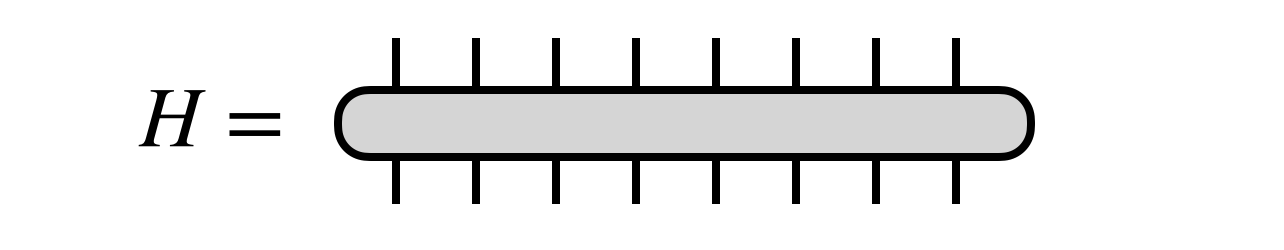}\label{fig: H_diagram}}
  \hfill
  \subfloat[MPO.]{\includegraphics[width= 0.5\columnwidth]{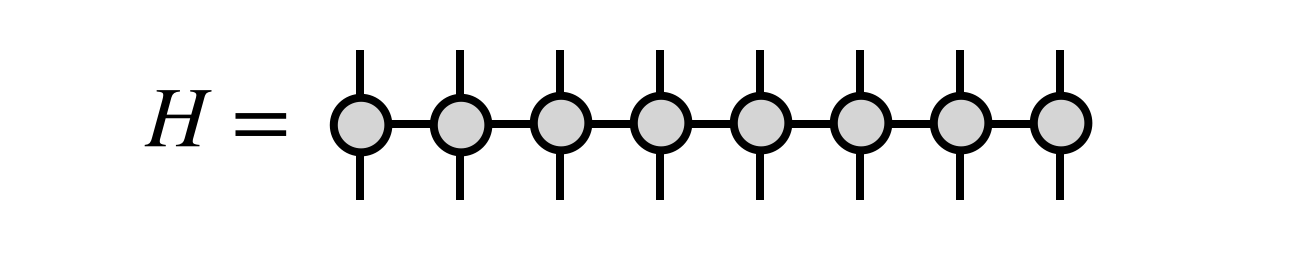}\label{fig: MPO}}
  \caption{(a) shows the representation of an operator as a $2N$-rank tensor, while (b) illustrates the equivalent TN but with an MPO structure.\protect\chapterthreefootnotemark}
\end{figure}

In the same way that we can represent states as MPS, there is a \textit{Matrix Product Operator} (MPO) structure suitable for representing an arbitrary operator with TN. A general operator can be written as
$$
H = \sum_{\{i\}}\sum_{\{i'\}} C_{i_1, \ldots i_L}^{i_1' \ldots i_L'}  |i_1 \ldots i_L \rangle \langle i_1' \ldots i_L' | \; \; \; .
$$
This structure is almost identical to the one we have seen with states but with more indices $i_k'$. The $C$ coefficients form a tensor with indices $i_1, \ldots i_L, i_1' \ldots i_L'$ as depicted in Fig. \ref{fig: H_diagram}.

\chapterthreefootnotetext

If we do a reshape to consider indices $i_k$, $i_k'$ as a single one, we can apply the same process as before to obtain the MPO structure
$$
H =\sum_{\{i\}}\sum_{\{i'\}} M_{i_1}^{i_1'} \ldots M_{i_N}^{i_N'} |i_1 \ldots i_L \rangle \langle i_1' \ldots i_L' | \; \; \; .
$$
See Fig. \ref{fig: MPO} for the graphical representation. We followed the same process that allowed us to create the MPS, so we know that this representation always exists. The only difference is that now our matrices have this double index, hence we can think of an MPO as matrices of matrices.

\begin{figure}[H]
    \centering
    \includegraphics[width= 0.4\columnwidth]{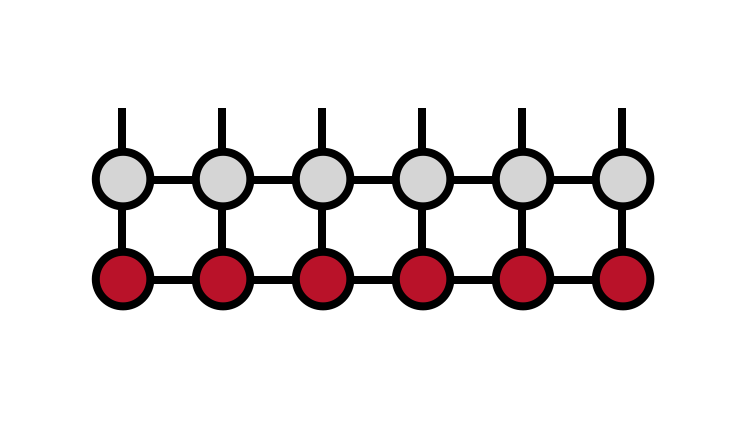}
    \caption{Contraction of an MPS $|\psi \rangle$ (red) with an MPO $H$ (grey), which is equivalent to the operation $H|\psi \rangle = |\psi' \rangle$. Notice that the result is a new MPS representing $|\psi' \rangle$. To determine the expected energy, we just have to insert the MPS $|\psi \rangle$ into the diagram with downward-pointing indices, as contracting the entire network would be equivalent to compute $\langle \psi | H | \psi \rangle$.\protect\chapterthreefootnotemark}
    \label{fig: finalMPS_retallada}
\end{figure}

Once we know how to represent an operator $H$ as an MPO, we can apply it to a state $|\psi \rangle$ represented as an MPS. Diagrammatically, it is as easy as contracting them through their physical indices as shown in Fig. \ref{fig: finalMPS_retallada}.

\chapterthreefootnotetext

\section{The DMRG algorithm} \label{sec: DMRG}

The \textit{two site Density Matrix Renormalization Group} (two site DMRG) algorithm aims to find the MPS that minimizes the energy of a given operator in the form of MPO. The idea is to variationally optimize the MPS tensors until a convergence criterion is met or until we reach a predefined limited number of iterations. 

\begin{figure}[H]
    \centering
    \includegraphics[width= 0.7\columnwidth]{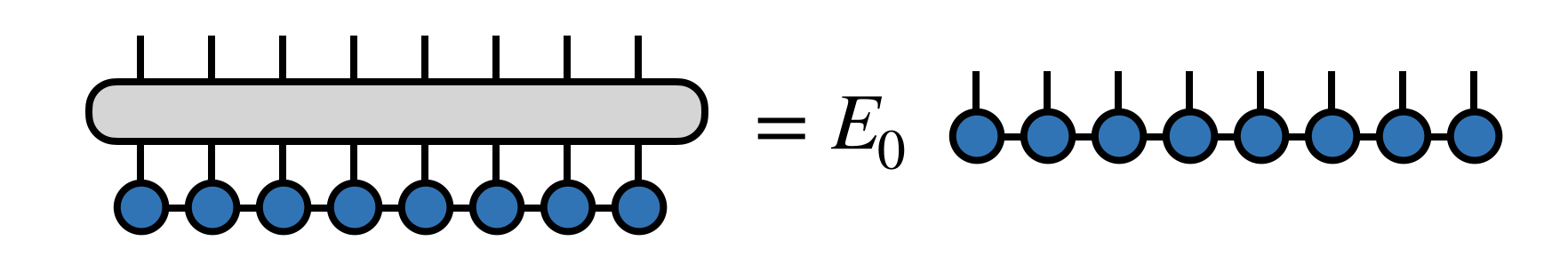}
    \caption{Equation $H|\psi\rangle = E_0 |\psi\rangle$ in tensor notation. The grey tensor describes the Hamiltonian operator, but we can also use its MPO form.\protect\chapterthreefootnotemark}
    \label{fig: H_eigenvector}
\end{figure}

The idea behind the DMRG is the following. In Equation \ref{eq:time-indep eq} we saw that energy was quantized and encoded in the eigenvalues of the Hamiltonian $H|\psi\rangle = E_0 |\psi\rangle$. This equation can be described with TN as depicted in Fig. \ref{fig: H_eigenvector}.

Applying the bra vector we obtain $\langle \psi | H | \psi \rangle  = \lambda \langle \psi | \psi \rangle$ where $\lambda$ is the eigenenergy. Therefore, the goal of the DMRG is to find $| \psi \rangle$ which gives us the minimum energy, i.e. the ground state: $$\langle \psi | H | \psi \rangle  / \langle \psi | \psi \rangle = \lambda = E_0 \; \; \; .$$


The issue is that this is a multilinear problem in the unknown variables (the MPS is formed by $N$ matrices) which would be very hard to compute. Instead, the two-site DMRG algorithm computes the minimization considering only two matrices at a time: we solve the problem for two sites of the MPS (\textit{optimization step}), we fix the solution obtained, and then we go to the next pair of consecutive tensors, and so on for a bunch of iterations until we converge to a solution of the whole problem. We solve the optimization mentioned for each pair of consecutive tensors from left to right and back again. This full iteration is called \textit{sweep}. 

Starting with a random MPS, we begin by computing its right-orthogonal form to allow an efficient contraction of the network. We can employ SVD for this purpose as explained in Section \ref{sec: MPS conversion}, and we can also use truncation to set an upper bound $\chi$ for the bond dimension.

\begin{figure}[htb]
    \centering
    \includegraphics[width= 0.45\columnwidth]{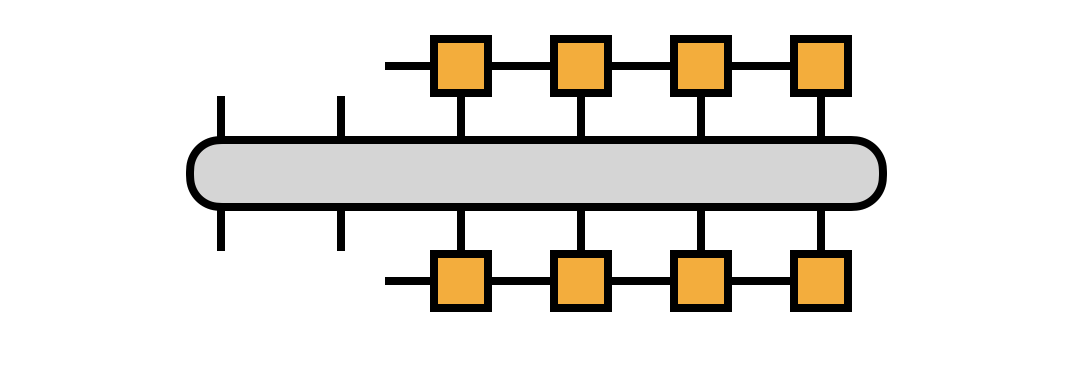}
    \caption{TN equivalent to $\langle \psi | H | \psi \rangle$ but without considering the first two sites. The goal is to determine the two MPS tensors that, when inserted there, yield the lowest energy.\protect\chapterthreefootnotemark}
    \label{fig: projected_H}
\end{figure}

\begin{figure}[ht]
  \centering
  \subfloat[Initial contraction of the DMRG.]{\includegraphics[width= 0.5\columnwidth]{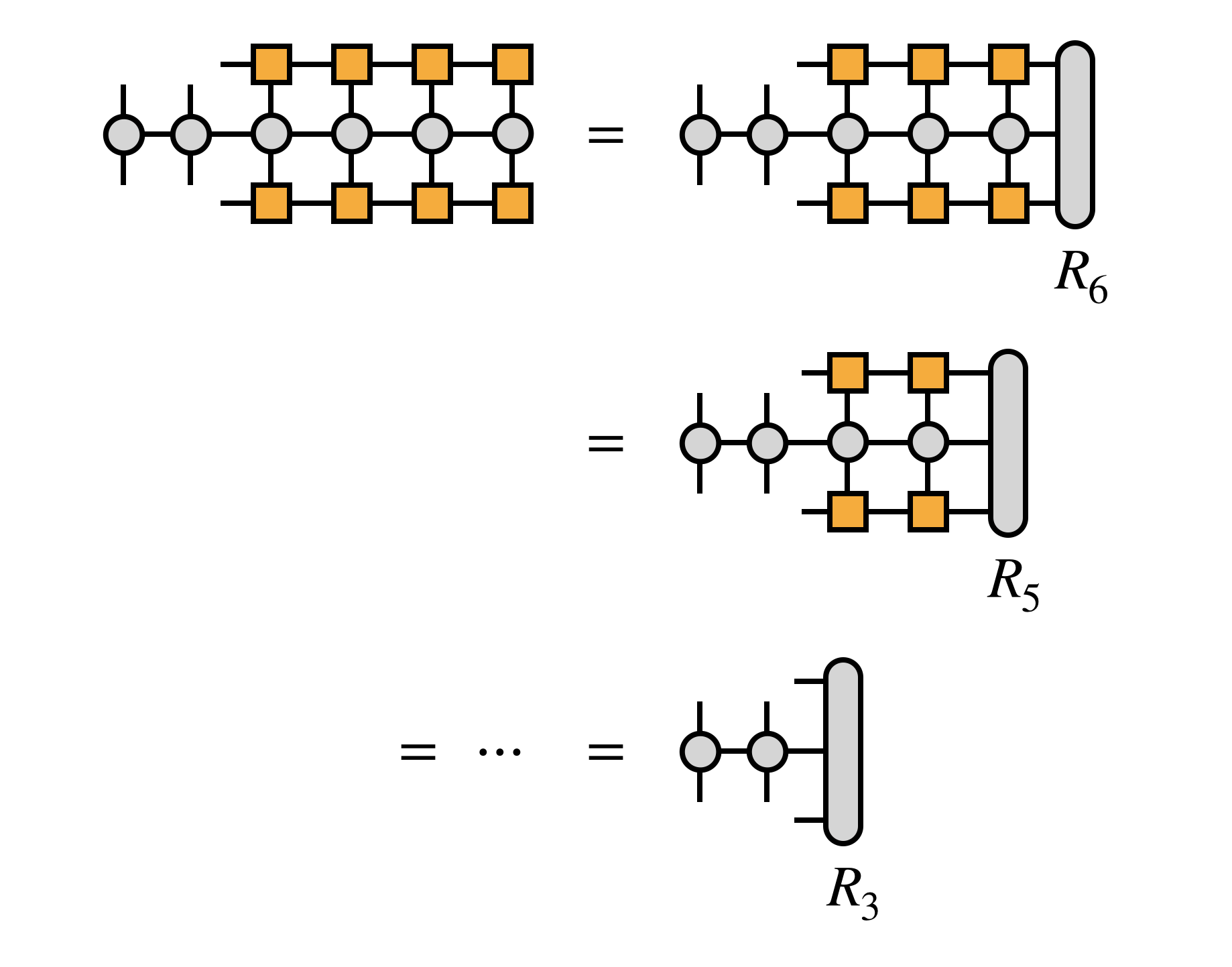}\label{fig: H_edge}}
  \hfill
  \subfloat[Optimal ordering for the initial contraction.]{\includegraphics[width= 0.5\columnwidth]{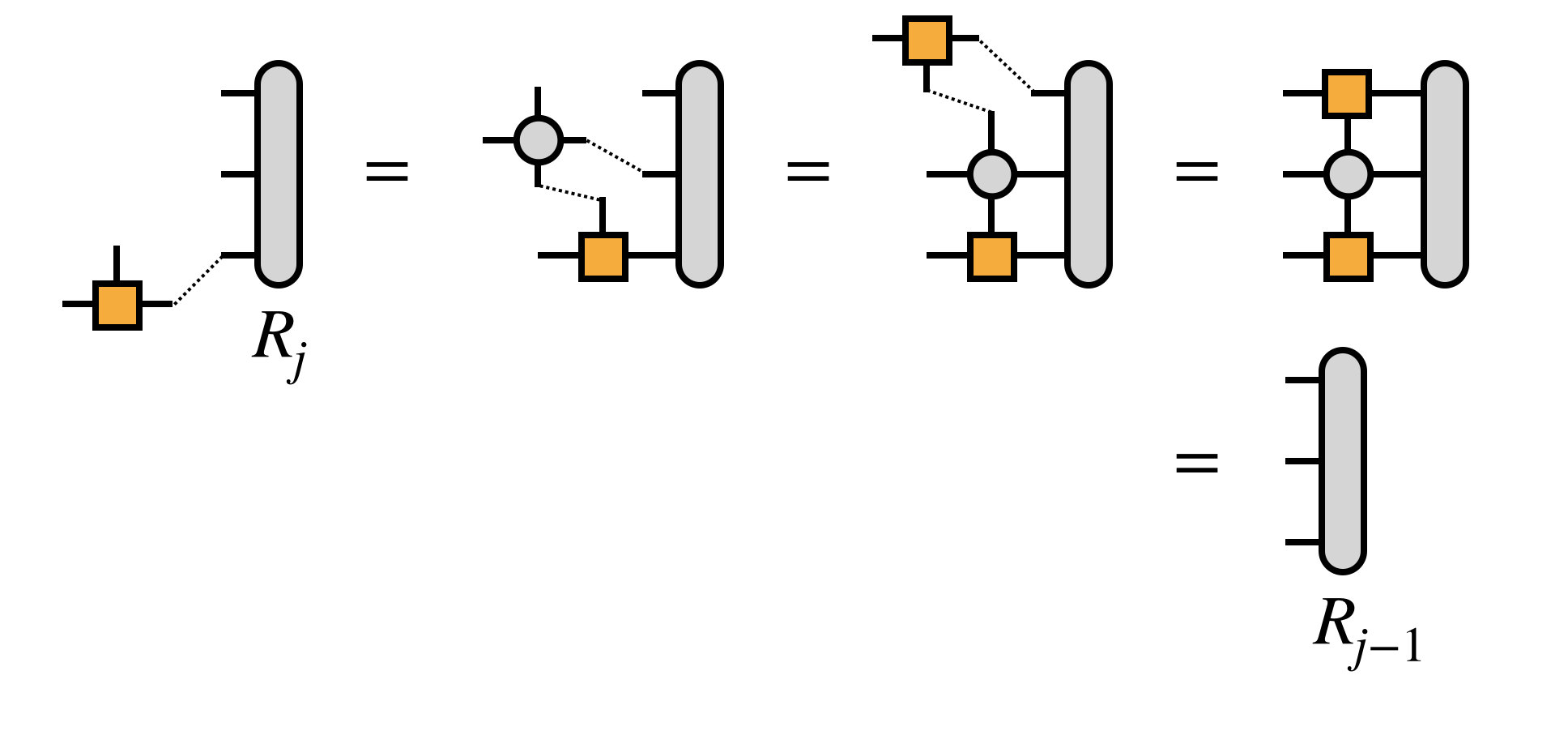}\label{fig: H_edge_ordering}}
  \caption{(a) illustrates how we should contract the initial TN leaving only the indices of the two sites to be optimized, while (b) describes the optimal sequence for performing the contractions to maximize computational efficiency.\protect\chapterthreefootnotemark}
\end{figure}

Then, we consider the representation of $\langle \psi | H | \psi \rangle$ without considering the first two sites as illustrated in Fig. \ref{fig: projected_H}. We can think of this as the matrix $H$ written in the basis $\{i_1, i_2, \alpha\}$ where $\alpha$ is the remaining open index (virtual index). We contract the whole network following the procedure of Fig. \ref{fig: H_edge}, in the order dictated by Fig \ref{fig: H_edge_ordering}.

\chapterthreefootnotetext

Through contraction and reshaping, we can interpret this TN as a matrix that describes a mapping from the space above to the space below ($\{i_1, i_2, \alpha\} \rightarrow \{i_1, i_2, \alpha\}$). We should store every $R_i$ obtained during the procedure shown in Fig. \ref{fig: H_edge} as they can be reused in the next iterations.

\begin{figure}[htb]
  \centering
  \subfloat[Contract the two tensors to be optimized.]{\includegraphics[width= 0.5\columnwidth]{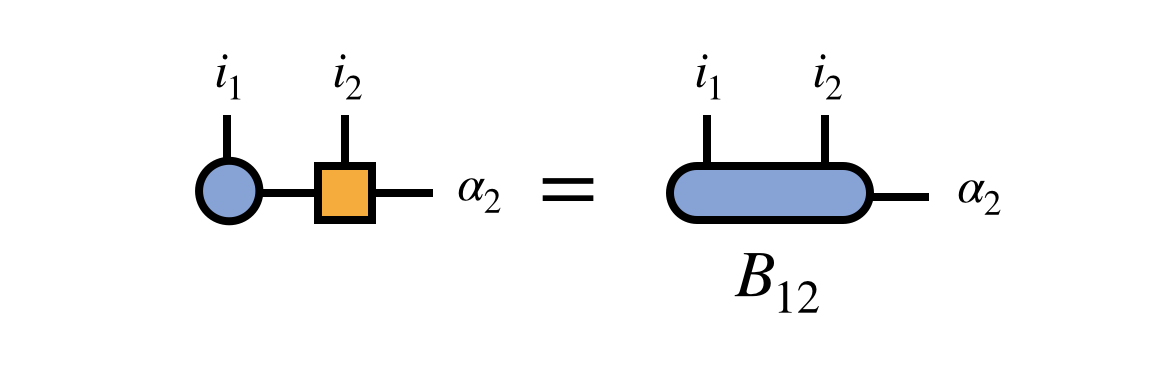}\label{fig: bond1}}
  \hfill
  \subfloat[Restore the MPS form after the optimization.]{\includegraphics[width= 0.5\columnwidth]{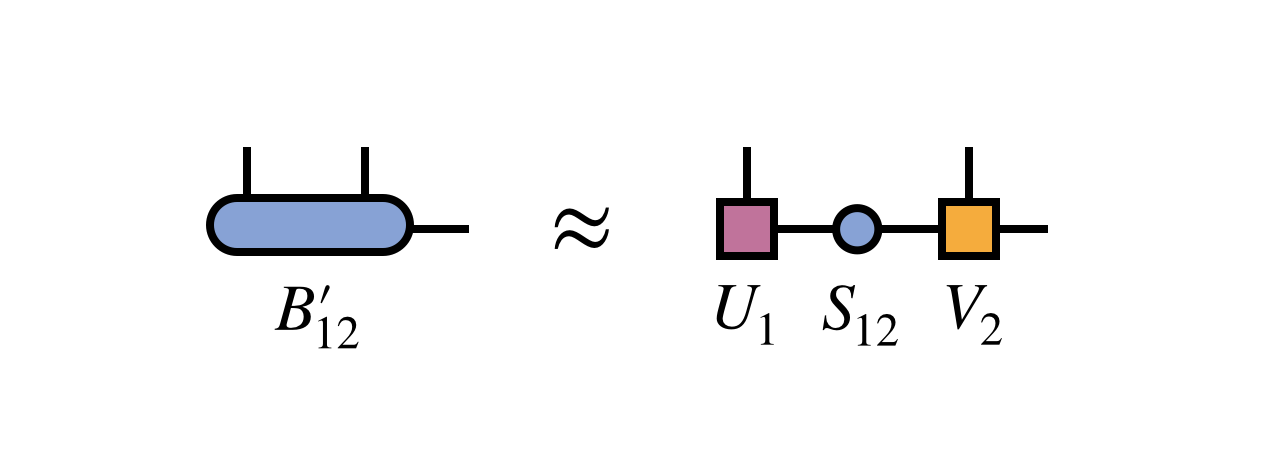}\label{fig: B12_svd}}
  \caption{(a) exhibits how the first two MPS tensors are contracted to start the optimization step, while (b) shows how to restore the MPS form after the optimization with the SVD (doing truncation if needed).\protect\chapterthreefootnotemark}
\end{figure}

\begin{figure}[ht]
  \centering
  \subfloat[Operator applied to the 1st pair.]{\includegraphics[width= 0.45\columnwidth]{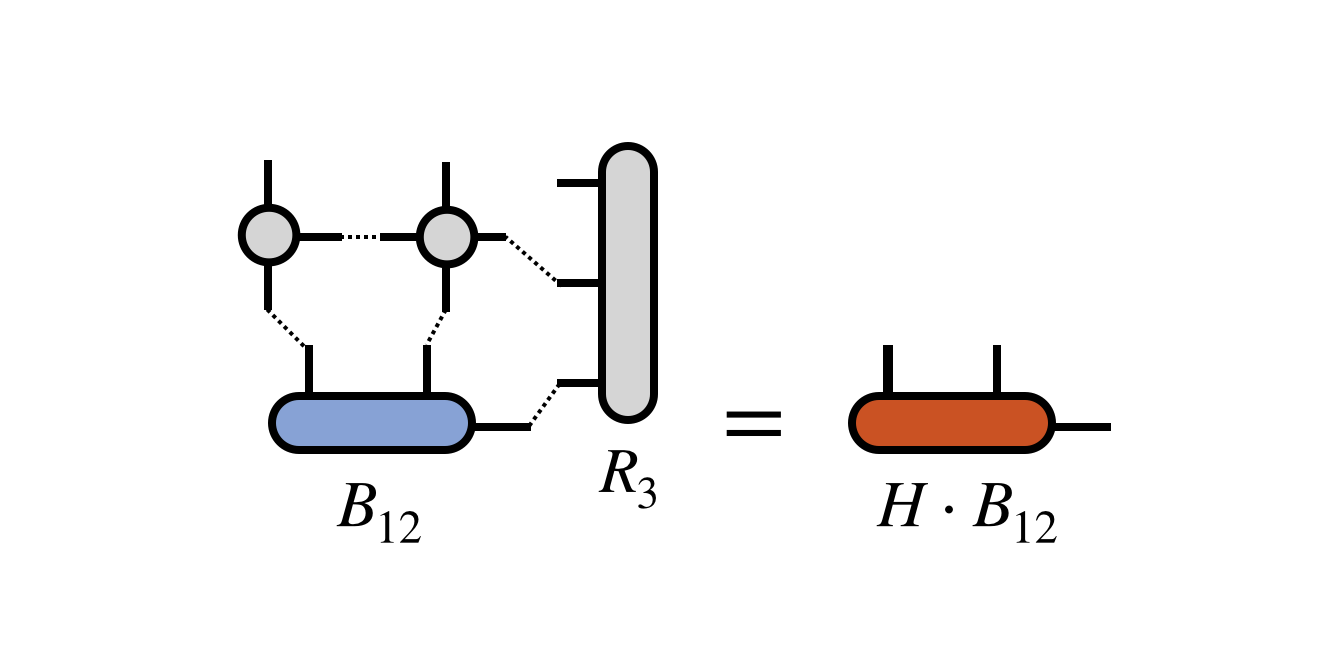}\label{fig: B12_H_mult}}
  \hfill
  \subfloat[Operator applied to the 2nd pair.]{\includegraphics[width= 0.55\columnwidth]{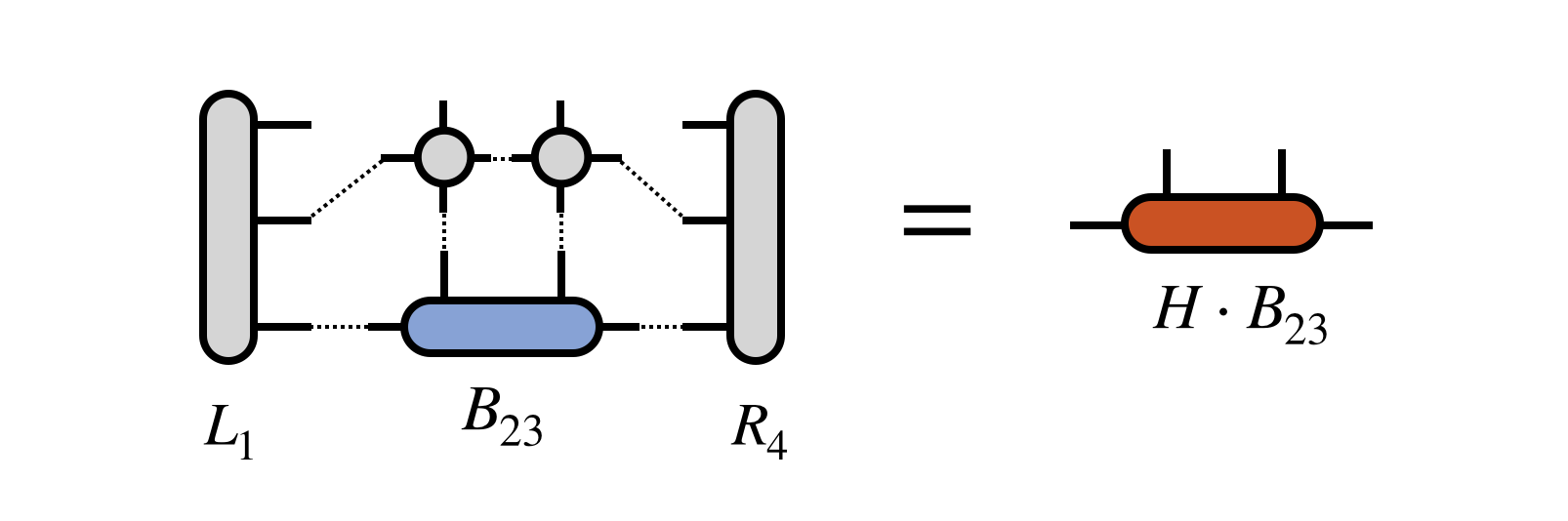}\label{fig: mult_H_B23}}
  \caption{(a) shows how to apply the operator to the leftmost pair of tensors of the MPS, while (b) illustrates the same process for the next pair of tensors. These are the operations that the Lanczos algorithm computes iteratively to approximate the lowest eigenvector of the operator.\protect\chapterthreefootnotemark}
\end{figure}

Next, we combine the first two MPS tensors by contracting over their shared bond index to obtain a tensor $B_{12}$ (Fig. \ref{fig: bond1}). $B_{12}$ can be seen as a vector, and together with the previously obtained local matrix $H$ (Fig. \ref{fig: B12_H_mult}) we have all the ingredients to execute the optimization step as an eigenvalue problem. It is generally solved using the \textit{Lanczos method} \cite{DMRG}, an iterative technique derived from the power method, that starts with an initial guess of the lowest eigenvector ($B_{12}$) and iteratively applies the matrix (local $H$) to find a better approximation of the lowest eigenvector $B_{12}'$. To get an idea of how this algorithm works see Appendix \ref{sec: power method}.

Finally, we compute the SVD of $B_{12}'$ to restore the MPS form (Fig. \ref{fig: B12_svd}). We absorb the diagonal matrix containing the singular values to the tensor at the right. We can also do a truncation to control the bond dimension and keep the efficiency of the operations. Notice that this new MPS will obey the left-orthogonal property since it is the SVD result. Now the MPS is in what is known as a \textit{mixed orthogonal form}, crucial for keeping the efficiency in the computations \cite{video_lecture_DMRG}.

\begin{figure}[htb]
  \centering
  \subfloat[Creation of the 1st left block.]{\includegraphics[width= 0.47\columnwidth]{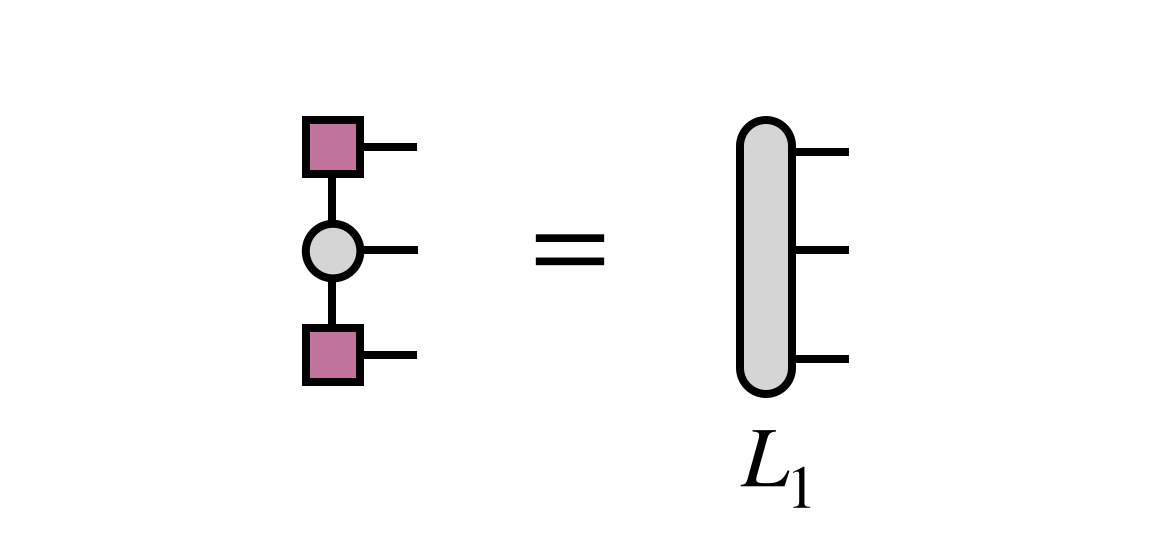}\label{fig: L1}}
  \hfill
  \subfloat[Creation of the 2nd left block.]{\includegraphics[width= 0.53\columnwidth]{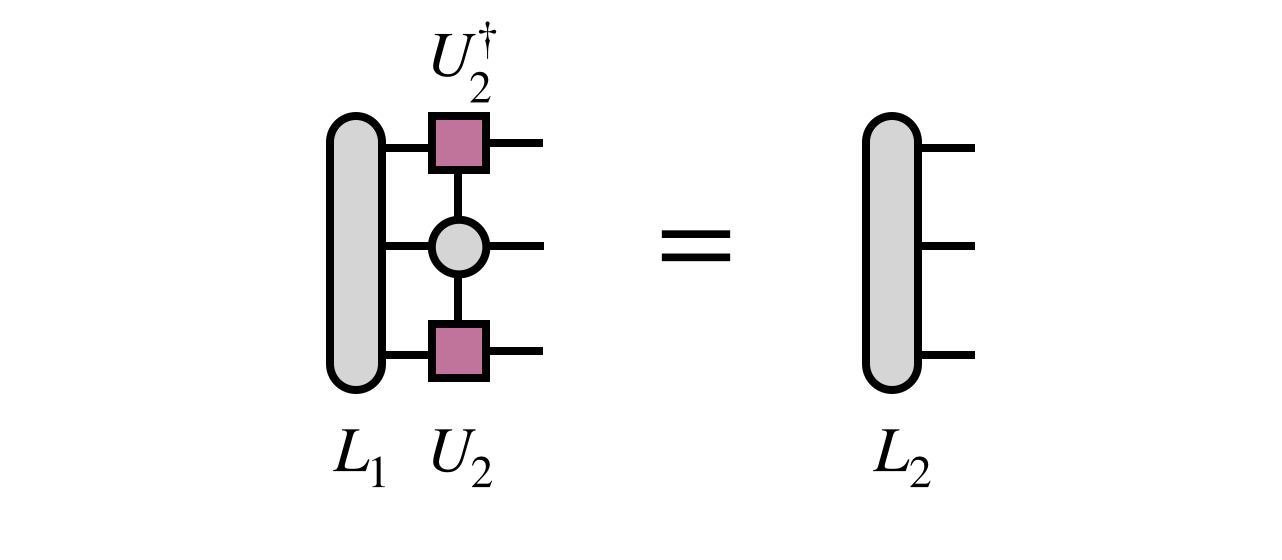}\label{fig: L2}}
  \caption{(a) depicts how to build the first left block after the first optimization to move into the next pair of tensors, and (b) illustrates the creation of the second left block needed for the third optimization.\protect\chapterthreefootnotemark}
\end{figure}
\chapterthreefootnotetext

After improving the first two MPS tensors, we go to the next consecutive pair and do the same optimization. We compute the $L_1$ tensor (Fig. \ref{fig: L1}), which can be done efficiently thanks now to the the left-orthogonality of the MPS, and then analogously solve the optimization step as before (Fig. \ref{fig: mult_H_B23}). These left blocks will contain the information of the already optimized tensors for the next optimizations. After the optimization step, we compute $L_2$ for the next iteration (Fig. \ref{fig: L2}) and so on until we reach the end of the chain. $R_i$ matrices do not need to be recomputed if they were saved in memory. To complete the sweep, we follow the same procedure from right to left.

We keep doing this until we reach a reasonably large number of sweeps, or until a convergence criterion is met. To monitor convergence, an easy approach is to look at the variance of the energy, which should disappear if you are in an eigenstate (remember Section \ref{sec: hamiltonians})
$$
\text{var(E)} = \langle \psi | H^2 | \psi \rangle  - \left( \langle \psi | H | \psi \rangle \right)^2 \; \; .
$$

However, take into account that we are seeking the state with the lowest energy but in an MPS form. Although MPS can represent any state, a limited bond dimension $\chi$ restricts this expressibility, which may prevent us from reaching the global minimum.

\chapterthreefootnotetext

\subsection{Estimation of the minimum gap}

We mentioned in Section \ref{sec: AQC intro} how important it is for AQC to study the minimum gap. However, we do not have many options to do so. For research purposes, the standard procedure is to diagonalize the Hamiltonian matrix and then compute the difference between the two lowest eigenvalues (which is already the solution to the minimization). However, diagonalizing a $2^N \times 2^N$ matrix is computationally very challenging \cite{TNBasicIntro}. This is the bottleneck if we want to study the annealing time required for different initial Hamiltonians, problem formulations, etc. 

To compute the minimum gap we need both the ground state and the first excited state. However, algorithms like QA and the DMRG just output the ground state. We can do a trick to obtain the first excited state \cite{kim2023variational} that consists in, once we have the ground state $|\psi\rangle$, run again the algorithm on the modified Hamiltonian of the form:
\begin{equation} \label{eq: ham with penalty}
    \mathcal{H}' = \mathcal{H} + w | \psi \rangle \langle \psi | \; \; ,
\end{equation}
where $w>0$ is the penalty weight that makes $|\psi\rangle$ no longer be the lowest energy state. $w$ needs to be at least larger than the minimum gap, but there is a trade-off because if it is larger than the maximum energy we lose precision, as we are making the spectrum wider. This modified Hamiltonian can be incorporated into the DMRG's optimization step without explicitly constructing a new Hamiltonian.

\sloppy This gives us a clear road map of how to employ the DMRG algorithm to estimate the minimum gap. Given the time-dependent Hamiltonian in Equation \ref{eq: AQC interpolation}, we run the DMRG algorithm for a discrete set of equally spaced time steps $\{ t_0, \ldots, t_n | \; t_i = i/n\}$ and get an estimate of the minimum energy levels $\{ \tilde{E_0}(t_0), \ldots, \tilde{E_0}(t_n) \}$. Then, we do the same process for the modified Hamiltonian in Equation \ref{eq: ham with penalty}, modifying the DMRG algorithm to minimize $$\langle \phi | \mathcal{H} | \phi \rangle + w \langle \phi | \psi \rangle \langle \psi | \phi \rangle  \; \; .$$ This will give us an estimate of the second energy level at each time step $\{ \tilde{E_1}(t_0), \ldots, \tilde{E_1}(t_n) \} $. Our minimum gap estimation would then be
\begin{equation} \label{eq: min gap estimate}
    \tilde{g}_{min} = \min_{i \in \{0, \ldots, n\}} (\tilde{E_1}(t_i) - \tilde{E_0}(t_i)) \; \; .
\end{equation}
\noindent This estimation could then be used to run QA or to study the energetic landscape, which is essential for understanding and improving QA performance and validating its results. It's important to note that DMRG can also find good solutions for the same optimization problem. In fact, $\tilde{E_0}(t_n)$ approximates the solution we are looking for.

%% file: chapters/C4_MPO.tex
\chapter{Construction of Matrix Product Operators using finite automata}
\label{sec: MPO}

One of the main challenges in applying DMRG is constructing the Matrix Product Operators (MPO) that represents the target Hamiltonian. As we explained in Section \ref{sec: hamiltonians}, a Hamiltonian is a $2^N \times 2^N$ Hermitian matrix, and the advantage of using an MPO is achieving a more efficient representation, where the number of parameters grows linearly instead of exponentially (see Section \ref{sec: TN representation of quantum systems}). Therefore, we require a method for constructing MPOs without constructing the entire Hamiltonian matrix, as doing so would incur a prohibitive computational expense.

Unfortunately, addressing this issue comprehensively for all scenarios can be challenging. There are some general methodologies, such as the one presented in Ref. \cite{Generic_construction}, but they often lack accessibility without a profound understanding of the underlying physics. This limitation significantly complicates the application of DMRG (along with other TN algorithms) to certain classes of Hamiltonians. In this section, we will provide an explanation on how to address this challenge by utilizing finite automata \cite{automataIntro, AppendixA}, a more comprehensive approach. Subsequently, we will construct the MPO for the Hamiltonian we are interested in: the annealing Hamiltonian of an Ising model.

\section{Construction of MPS and MPO}

An MPO can be seen as a long sequence of matrices of matrices (rank-4 tensors), which makes building and visualizing them quite challenging. However, in Ref. \cite{automataIntro} a clear visualization of the matrix product is presented using directed graphs, establishing a link between MPOs and automata theory.

The mapping from automata to matrix products can serve for the construction of both MPO and MPS. The only difference is that instead of working with states (0 or 1) we work with operators.

\subsubsection{Construction of MPSs}

\begin{figure}[H]
    \begin{center}
    \includegraphics[width= 0.5\columnwidth]{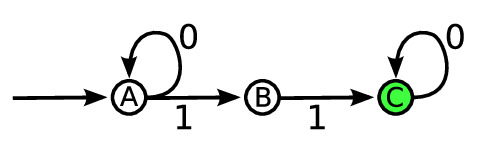}
    \end{center}
    \caption{Automaton that accepts binary strings with exactly two consecutive 1s. Image from Ref. \cite{automataIntro}.}
    \label{fig: automata}
\end{figure}

Let's illustrate the procedure with an example: let's say we want to build the MPS that represents an equal superposition of the set of states with two neighboring 1's, and the rest of qubits in the 0 state
$$
11000\ldots + 01100\ldots + 00110\ldots + \ldots
$$
the summands of this expression form the language of the finite automaton shown in Fig. \ref{fig: automata}.

\begin{figure}[H]
    \centering
    \includegraphics[width= \columnwidth]{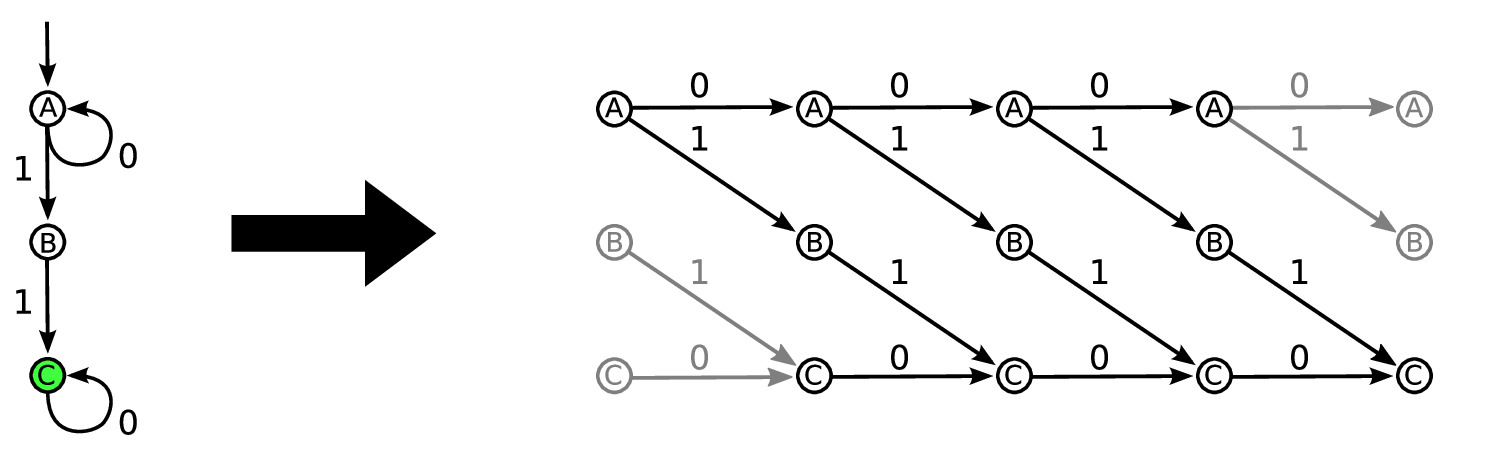}
    \caption{Mapping from finite-state automaton to MPS diagram. Image from Ref. \cite{automataIntro}.}
    \label{fig: MPO diagram}    
\end{figure}

We can transform this automaton into an MPS diagram by doing the transformation shown in Fig. \ref{fig: MPO diagram}. The MPS diagram is an exact representation of how the MPS should be. To illustrate this, let's show the equivalent MPS for a system of 4 particles, representing states 0 as $\downarrow$ and 1 as $\uparrow$:
$$
\begin{bmatrix}
\downarrow & \uparrow & 0
\end{bmatrix}
\cdot
\begin{bmatrix}
\downarrow & \uparrow & 0 \\
0 & 0 & \uparrow \\
0 & 0 & \downarrow 
\end{bmatrix}
\cdot
\begin{bmatrix}
\downarrow & \uparrow & 0 \\
0 & 0 & \uparrow \\
0 & 0 & \downarrow 
\end{bmatrix}
\cdot
\begin{bmatrix}
0 \\
\uparrow \\
\downarrow
\end{bmatrix}
= \; \downarrow\downarrow\uparrow\uparrow + \downarrow\uparrow\uparrow\downarrow + \uparrow\uparrow\downarrow\downarrow \; \; .
$$

\begin{figure}[h]
  \centering
  \hfill
  \hfill
  \begin{minipage}[c]{0.3\textwidth}
    \centering
    \includegraphics[width=\linewidth]{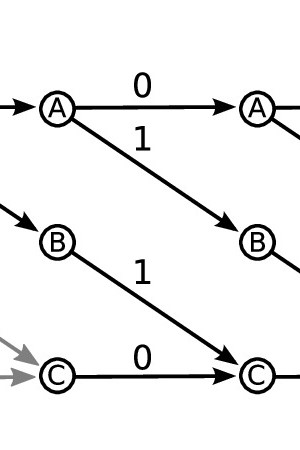}
  \end{minipage}
  \hfill
  \begin{minipage}[c]{0.5\textwidth}
    \centering
    \Huge 
    $\begin{bmatrix}
    \downarrow & \uparrow & 0 \\
    0 & 0 & \uparrow \\
    0 & 0 & \downarrow 
    \end{bmatrix}$
  \end{minipage}
  \caption{The arrows in each column denote the elements of the corresponding matrix, with nodes $A$, $B$, and $C$ representing indices 1, 2, and 3, respectively. Consequently, an arrow connecting index $i$ to index $j$ represents the element $ij$ in the matrix, and any omitted arrows are filled with $0$. For example, an edge in the 3rd column from node $A$ to node $B$ with a value of 1 indicates that in the 3rd matrix $M_3$ of the MPO, we have $m_{1,2} = \uparrow$.}
  \label{fig: mypaper-diagram-13_cut}
\end{figure}

To derive this MPS from the MPS diagram in Fig. \ref{fig: MPO diagram} we have to read the diagram from left to right. Note that nodes $A$, $B$, and $C$ appear repeatedly, forming columns. There are five of these columns, connected by directed edges that always link one column to the next in the sequence. This results in four columns of arrows, representing the four matrices of the MPS. Fig. \ref{fig: mypaper-diagram-13_cut} displays how the mapping from columns to matrices is done. The first and last matrices are simply the first row and the last column, respectively, ensuring that the final product is a $1 \times 1$ matrix.

The matrix at position $k$ defines the state of the $k$-th particle. In this case, the configuration follows the same pattern for all particles, leading to identical matrices throughout the product. This matrix is typically called the \textit{bulk matrix}.

The diagram serves as an effective representation of the matrix product of multiple matrices. Each summand in the final expression corresponds to a path from the leftmost node $A$ to the rightmost node $C$, concatenating all the elements encountered along the way. Thinking in terms of paths within a diagram provides a useful framework for creating matrix products from scratch. Another advantage of this method is that we can use the tools and techniques from automata theory to study and construct more complex matrix products, like union, concatenation, and intersection operations of their languages.

\subsubsection{Construction of MPOs}

\begin{figure}[H]
    \begin{center}
    \includegraphics[width= 0.7\columnwidth]{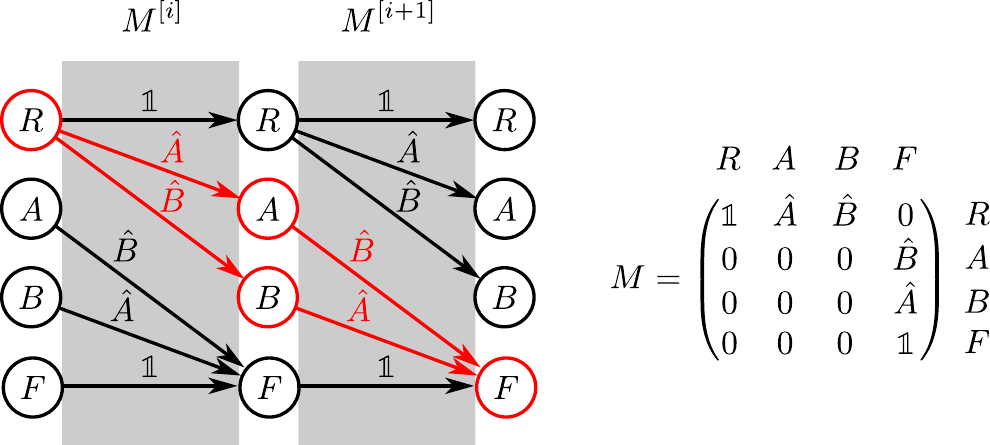}
    \end{center}
    \caption{MPS diagram and corresponding MPO matrix of the operator in Equation \ref{eq: MPO example}. Image from Ref. \cite{cylinder}.}
    \label{fig: cylinder MPO}
\end{figure}

The same procedure applies to the creation of an MPO. The only difference is that the arrows represent operators (rank-2 tensors) instead of states ($\uparrow$ or $\downarrow$). Let's say we want to build the operator
\begin{equation} \label{eq: MPO example}
    O = \sum_i \left( \hat{A}_{i}\hat{B}_{i+1} + \hat{B}_{i}\hat{A}_{i+1} \right)
\end{equation}
\noindent where $\hat{A}_i, \hat{B}_i$ are any operators $\hat{A},\hat{B}$ applied to the $i$-th particle. The Fig. \ref{fig: cylinder MPO} greatly illustrates the diagram and corresponding bulk tensor of the MPO. Indices $R$, $A$, $B$, and $F$ are the virtual indices of the MPO, so the bond dimension in this case is 4. The physical indices would be the ones given by matrices $\hat{A}$ and $\hat{B}$.

As a final remark, this procedure can be difficult to apply for long-ranged, generic Hamiltonians, such as the ones in chemistry. There are other approaches to tackle these more complex interactions as in Ref. \cite{Generic_construction}. Nonetheless, this reasoning framework is very convenient for the annealing Hamiltonian, which is our current focus of interest.

\section{MPO for the annealing Hamiltonian of an Ising model}
\label{sec: MPO for Annealing ham}

When MPOs get more complex, visualizing them via diagrams can be complicated. That is why in Ref. \cite{AppendixA} the authors use tables that contain the same information. For example, the MPS diagram in Figure \ref{fig: MPO diagram} could be represented by the table

$$
\begin{array}{c|ccc}
\text { rule-number } & \text { (left, right)-input } & & \text { output } \\
\hline 1 & (1,1) & \rightarrow & \downarrow \\
2 & (1, 2) & \rightarrow & \uparrow \\
3 & (2, 3) & \rightarrow & \uparrow \\
4 & (3, 3) & \rightarrow & \downarrow
\end{array}
$$

The table represents a column of arrows in the diagram. Rules tell us how to fill a concrete position in the final matrix $(M)_{i,j}$, the same information that the arrow was giving us in the diagram. Left and right indices refer to the nodes that the arrow is connecting, i.e. they fix a row $i$ and column $j$ of $M$, and \textit{output} is the element we should put in that position. For instance, the 3rd rule is telling us that $m_{2,3} = \uparrow$.

The nomenclature of \textit{(left, right)-input} and \textit{output} is used to enforce the idea that this is indeed an automaton. The automaton in Fig. \ref{fig: automata} had three states, as the number of possible left/right inputs, and four transitions, as the number of rules in our table. For instance, the 3rd rule describes the transition from state 2 to state 3, which happens after reading the second consecutive 1.

Using this notation, Appendix A of Ref. \cite{AppendixA} provides the tables for constructing the MPO of the Hamiltonian
$$
H = \sum^N_{i=1}X'_i + \sum_{k<l}c_{kl}Z_k \otimes Z_l
$$
and the proof that it has the lowest bond dimension possible. Nonetheless, it does not matter if we use the Pauli X or any other operator, as we can plug any operator in the corresponding positions of the MPO matrices.

Now, if we do the following change of variables
\begin{align*}
& X'_i = -(1-s) \cdot X_i+s \cdot h_i Z_i \\
& c_{kl} = s \cdot J_{kl} \; \; \; , J_{kl} \in \mathbb{R}
\end{align*}
\noindent we obtain
$$
H = \sum^N_{i=1}\left(-(1-s) \cdot X_i+s \cdot h_i Z_i\right) + \sum_{k<l}s \cdot J_{kl} \cdot Z_k \otimes Z_l
$$
which is the annealing Hamiltonian we proposed in Section \ref{sec: QA} for the Ising model \cite{combarro2023practical}

\begin{equation} \label{eq: annealing ham}
    H=-(1-s) \sum_i X_i+s\left(\sum_i h_i Z_i+\sum_{i j} J_{i j} Z_i Z_j\right) \; \; .
\end{equation}

\subsubsection{Our contribution}

We took the tables from Appendix A of Ref. \cite{AppendixA} and applied the specified change of variables to derive the MPO for the annealing Hamiltonian in Equation \ref{eq: annealing ham}. Additionally, the diagram visualization enabled us to improve their implementation through the following modifications:

\begin{itemize}
    \item They utilized auxiliary matrices to encode the coefficients $h_i$ and $J_{ij}$ for clarity. We extracted the underlying pattern without relying on these auxiliary matrices to achieve a more efficient representation, as it saves storage and computation by removing the need for $N-1$ matrices and matrix products.
    \item There are two errors in their implementation
    \begin{itemize}
        \item Bond dimension of $A^{[k]}$ in table V should be $D = \min(k+2, N-k+3)$
        \item The 3rd rule of table VI should have (left, right)-input $(m, k+2)$
    \end{itemize}
    
    \item They assumed an even number of particles. We extended the code to handle any given $N$. To achieve this, we introduce a variable $a$ in our formulation, where $a=2$ if $N$ is even, and $a=3$ otherwise. You may refer to Appendix \ref{sec: Tables MPO} for further details.
\end{itemize}

The final tables describing the matrices $M_k$ in our MPO implementation for the annealing Hamiltonian are located in Appendix \ref{sec: Tables MPO}. While we did not find an MPO for this Hamiltonian in the existing literature, we validated its correctness through two methods. First, for small instances, we contracted the MPO and confirmed it matched the symbolic expression in Equation \ref{eq: annealing ham}. For larger instances, we diagonalized both the real Hamiltonian matrix and the matrix obtained from contracting the MPO. Both matrices had the same spectral decomposition, particularly the same ground state.

%% file: chapters/C5_QKP.tex
\chapter{Application: solving the quadratic knapsack problem}

The knapsack problem is a powerful tool for computer scientists, as it acts as a fundamental framework for reasoning about a vast collection of problems \cite{glover2019tutorial}. Despite its basic formulation, many problems in industry can be modeled as knapsack problems \cite{dp_QKP}. Therefore, breakthroughs in solving the knapsack problem hold the potential to unlock solutions for many other practical cases.

The problem states the following \cite{Ising_formulations}: you are given a knapsack, like a backpack, with a limited weight capacity, and a set of items, each with its own weight and value. Your goal is to fill the knapsack with the most valuable selection of items possible without going over the weight limit.

To add a little bit of spice, and exploit the capacity of the Quadratic Unconstrained Binary Optimization (QUBO) formulation, we will extend the problem by giving extra profit if certain pairs of items are selected together. For instance, imagine that there is an item 'napkin' and some greasy food items. With this extension, we could give a very low value to napkin, as it alone has no value, but give an extra bonus if some greasy food appears in the solution, as it becomes an essential item in that case. This extension is known as the quadratic knapsack problem (QKP) \cite{dp_QKP, glover2019tutorial}, and it has many applications in finance, logistics, telecommunications, etc.

The chapter will follow this structure: we will start by introducing the problem's general formulation and encoding the inequality constraints using \textit{unbalanced penalization} \cite{montanezbarrera2023unbalanced} converting the problem into a QUBO to leverage quantum computing capabilities. Subsequently, we will generate the corresponding MPO and utilize the DMRG algorithm to examine adiabatic evolution. Using this insight, we will define a tailored annealing schedule for quantum annealing. Finally, we will present results for large QKP instances, comparing them with our dynamic programming implementation, which is explained in Appendix \ref{sec: DP algorithm}.

\section{General formulation}

Let $W$ be the weight capacity of our knapsack, and $N$ be the number of available items to put in the bag. Each item has a weight $w_i$ and a value, but instead of representing it as $v_i$, we will use $v_{ii}$, so that we don't need extra notation to encode the extra profits of the QKP: $v_{ij}$ will be the extra profit obtained if items $i$, $j$ are both in the solution, for $i,j=1, ..., N$. To save space in storing $v_{ij}$, we can impose $i\ge j$ so that it fits in a triangular matrix

Let $x_i$ be the decision variables, which take values $1$ if we decide to put the $i$-th item in the bag, 0 otherwise.

The total weight of our solution is given by
$$
\mathcal{W}=\sum_{i=1}^N w_i x_i \; \; ,
$$
and the total value is
$$
\mathcal{V}=\sum_{i=1}^N \sum_{j=1}^i v_{ij} x_ix_j \; \; .
$$
Our goal is to maximize $\mathcal{V}$ subject to the constraint that $\mathcal{W}\leq W$.

\subsubsection{Imposing inequality constraints for QUBO}
\label{sec: ineq constraints}

The standard and most common approach to encode inequality constraints into a QUBO formulation is to transform them into equalities by using what is called \textit{slack variables} \cite{glover2019tutorial}. However, slack variables can significantly increase the number of variables needed, which is a bad trade-off in the current Noisy Intermediate-Scale Quantum (NISQ) era \cite{9251243} as quantum devices have limited amounts of qubits available. Therefore, we have opted to use a new technique presented in Ref. \cite{montanezbarrera2023unbalanced} that does not require any additional variable. Instead, it uses an \textit{unbalanced penalization function} that gives larger penalizations when the inequality constraint is not achieved. This function is an exponential function, approximated by its corresponding Taylor expansion.

Given an inequality constraint of the form
$$
\sum_{i} l_i x_i \le B, \; l_i \in \mathbb{Z} \; \; ,
$$
in our case $B = W$ and $l_i = w_i$, we can rewrite it as
$$
h(x) = B - \sum_{i} l_i x_i \ge 0 \; \; .
$$
We need to introduce this term in our QUBO formulation, so that non-allowed values get large penalizations, but without penalizing solution candidates. Therefore, the shape of the exponential decay curve, $f(x) = e^{-h(x)}$ is what we need, because:
\begin{itemize}
    \item If $h(x) \ge 0$ $\implies$ $f(x) \in (0, 1]$ $\implies$ almost no contribution to the total cost
    \item If $h(x) < 0$ $\implies$ $f(x)$ grows exponentially
\end{itemize}
However, to encode an exponential function in a QUBO formulation we need to consider its Taylor expansion to at most the quadratic term
$$
e^{-h(x)} \approx 1 - h(x) + \frac{1}{2}h(x)^2 \; \; .
$$
Moreover, in Ref. \cite{montanezbarrera2023unbalanced} they introduce some tunable parameters and compute the most appropriate values for the KP problem with the Nelder-Mead optimization method. They end up with the expression
$$
g(h(x)) = - 0.9603 \cdot h(x) + 0.0371 \cdot h(x)^2 \; \; .
$$
They compute these values for a classical KP. In our experiments, we obtained overweighted solutions, above the knapsack capacity. To solve this issue, we added a multiplicative factor $\lambda g(h(x))$ that controls the slope of the parabola. 

In Fig. \ref{fig: unbalanced penalization} we can appreciate the three effects of this $\lambda$ factor:

\begin{figure}[htb]
    \begin{center}
    \includegraphics[width= 0.8\columnwidth]{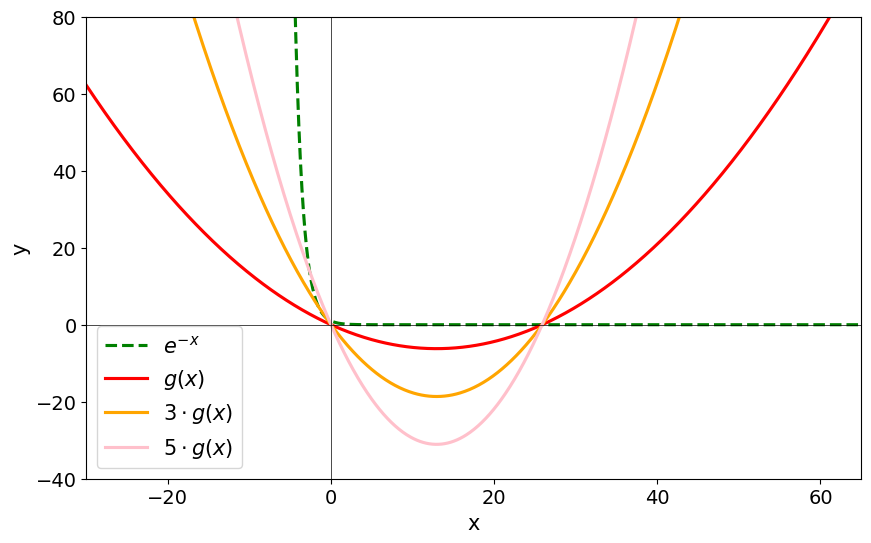}
    \end{center}
    \caption{The x-axis represents $h(x)$, which means that values at the left-hand side of the y-axis should exponentially increase to penalize non-valid solutions. In green the ideal function to encode the constraint $e^{-h(x)}$. The red plot represents $g(h(x))$, proposed in Ref. \cite{montanezbarrera2023unbalanced}. Orange and pink plots show the effect of our introduced parameter $\lambda$ for $\lambda=3,5$.}
    \label{fig: unbalanced penalization}
\end{figure}

\begin{itemize}
    \item[(a)] The bigger the $\lambda$, the greater the penalization is for overweight candidates, as the function gets closer to the exponential in the second quadrant.
    \item[(b)] Candidates with a weight between 0 and $25.88$ (approximately) below the limit (cuts with the x-axis) get a small positive reward, which gets bigger as we increase $\lambda$. This is not a good thing but sounds like a good heuristic, as it is aligned with the intuition that the more we fill our knapsack, the more value we can carry. 
    \item[(c)] Candidates with a weight below $W - 25.88$ (approximately) are unfairly penalized, and increasing $\lambda$ makes the problem worst.
\end{itemize}

Therefore, picking the right $\lambda$ value involves making a trade-off. As our problem was that the constraint was not being satisfied, we had to increase the value of $\lambda$, and we got incredibly better results. In future work, conducting a more complete hyperparameter tuning would be nice to find a better choice for $\lambda$.

\subsubsection{Computation of the MPO}

The final cost function that we want to minimize is 
\begin{align}
    cost &= - \mathcal{V} + \lambda \left[- 0.9603 \cdot h(x) + 0.0371 \cdot h(x)^2 \right] \label{eq: QUBO cost} \\
    &= - \sum_{i=1}^N \sum_{j=1}^i v_{ij} x_ix_j + \lambda \left[ - 0.9603 \cdot \left(W - \sum_{i=1}^N w_i x_i\right) + 0.0371 \cdot \left(W - \sum_{i=1}^N w_i x_i\right)^2 \right] \; \; . \notag
\end{align}
This can be directly mapped into an Ising Hamiltonian (Equation \ref{eq: Ising ham}) just by applying the procedure explained in Section \ref{sec: QUBO intro}. However, we must take into account that the dimod library uses the change of variables
$$
x = \left( \frac{1 + s}{2}\right)
$$
Which maps $x=0 \iff s=-1$ and $x=1 \iff s=1$. The only implication of this is that we have to flip the bit values of the final solution.

Once we have the Ising Hamiltonian, we take the obtained coefficients and apply the rules in the tables of Appendix \ref{sec: Tables MPO} to obtain the MPO of the corresponding annealing Hamiltonian. We have all the ingredients to run both QA and the DMRG.

\section{Simulated annealing approach}

\begin{figure}[htb]
    \centering
    \includegraphics[width= \columnwidth]{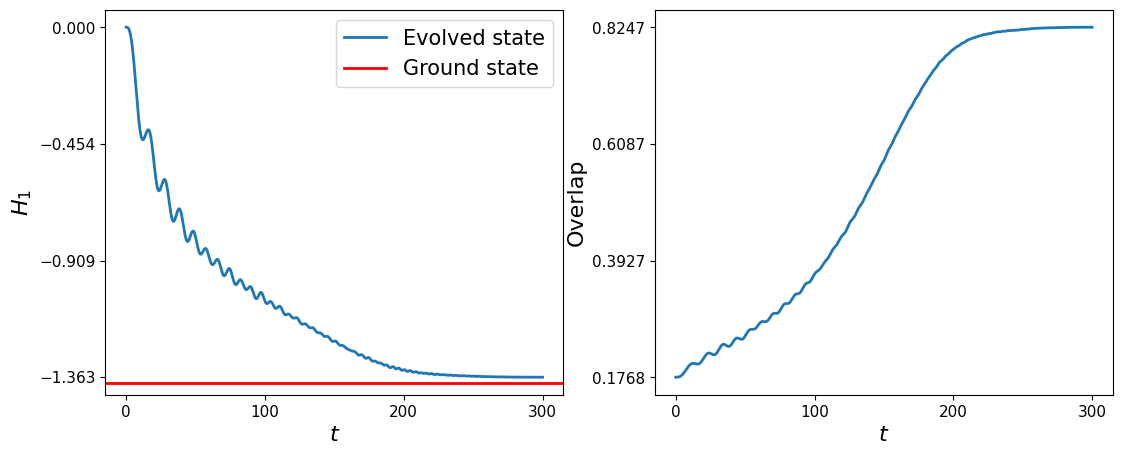}
    \caption{Plot of the simulated adiabatic evolution performed with Qibo for a QKP instance with 10 items. The left plot displays the evolution of the energy in blue, and it closely approaches the actual ground state energy (in red) computed using exact diagonalization. The right plot shows the overlap between the state at each time step and the exact solution state (the eigenvector with the minimum eigenvalue). A final overlap greater than 0.8 indicates that our quantum state will most probably collapse to the target state when measured. The solution obtained matches the one derived using brute force.}
    \label{fig: sim QA N10}
\end{figure}

Unfortunately, we do not have access to a quantum annealer. Instead, we used the Qibo library \cite{qibo}, which provides a full-stack API for quantum simulation and quantum hardware control. This library includes a method that takes an initial and target Hamiltonian, along with an annealing schedule time ($s(t)$ in Section \ref{sec: AQC intro}), and runs a simulated adiabatic evolution. The simulation internally handles the $2^N \times 2^N$ matrix Hamiltonian, which limits us to QKP instances with no more than approximately 10 to 12 items. The good thing is that for these instances, we can also perform exact diagonalization of the matrices to compare the results. See Fig. \ref{fig: sim QA N10} to see the annealing evolution for a QKP instance of 10 items.

\subsubsection{Computing the minimum gap using DMRG}

\begin{figure}[htb]
  \centering
  \subfloat[Simmulated annealing run with DMRG.]{\includegraphics[width= 0.5\columnwidth]{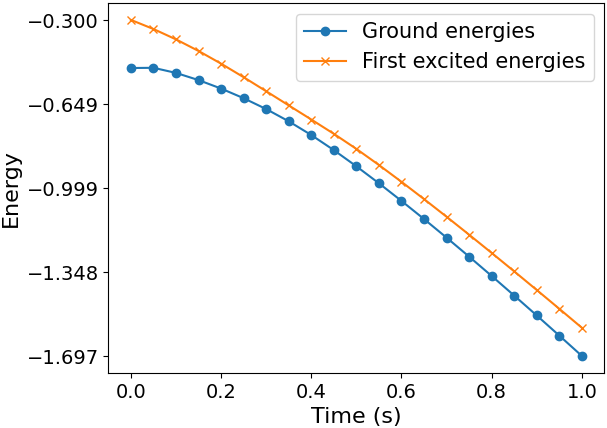}\label{fig: annealing_run}}
  \hfill
  \subfloat[Test the correctness of the gaps obtained.]{\includegraphics[width= 0.5\columnwidth]{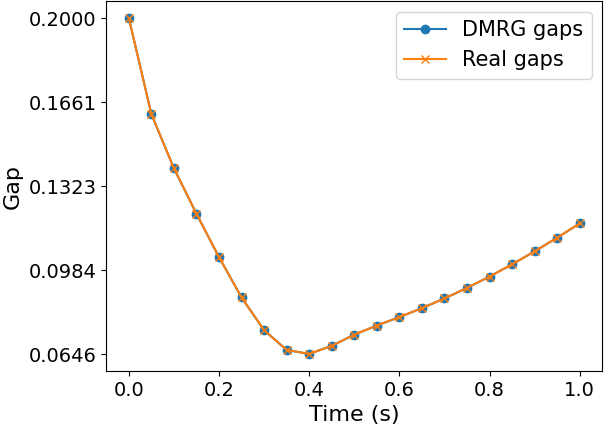}\label{fig: gaps}}
  \caption{Both plots correspond to the same instance of the QKP with 5 items. (a) shows the simulated annealing evolution using DMRG taking 20 equidistant time steps, while (b) compares the gaps obtained with the exact gaps computed using exact diagonalization of the Hamiltonian matrix, to showcase that the computed evolution is correct.}
  \label{fig: DMRG gaps}
\end{figure}

We implemented the procedure explained in Section \ref{sec: DMRG} to approximate the gaps of the annealing evolution. An overview of how the procedure works is provided in Algorithm \ref{alg: DMRG}, and an example of the output is shown in Fig. \ref{fig: DMRG gaps}.
\begin{algorithm}
\caption{Scan the gap of an annealing run with DMRG}\label{alg: DMRG}
\begin{algorithmic}
\Require $MPO(s)$ returns the annealing Hamiltonian at time step $s$
\Require $DMRG(M, | \psi \rangle)$ runs the DMRG for MPO $M$ and penalizing state $| \psi \rangle$
\State Input the number of annealing steps $T$
\State $T \gets T-1$
\State Initialize $G(k) \gets 0$ for $k=0, \ldots T$ \Comment{stores the gaps at each time step}

\For{$s=0, \;  1/T,\; 2/T, \ldots,\; (T-1)/T,\; 1$}
    \State $M \gets MPO(s)$
    \State $gs\_energy, | \psi \rangle \gets DMRG(M, \emptyset)$ \Comment{computes the ground state}
    \State $exc\_energy, | \phi \rangle \gets DMRG(M, | \psi \rangle)$ \Comment{computes the 1st excited state}
    \State $G(k) \gets exc\_energy - gs\_energy$
\EndFor

\State Output $G(k)$
\end{algorithmic}
\end{algorithm}

We want to highlight a technique that significantly improved the accuracy of our results. Instead of using a random initialization for the initial MPS in the second round of the DMRG to compute the first excited state, we started with the MPS equivalent to the $|0\ldots0\rangle$ state. We reasoned that starting the optimization from a state orthogonal to the ground state might aid in convergence, as this is a property the solution should possess. In future work, we would like to explore the efficiency of this initialization compared to the random one.

The $|0\ldots0\rangle$ state should be orthogonal to the ground state, as we expect our solution to have at least one item. This orthogonality occurs only at the final time step when the annealing Hamiltonian equals the target Hamiltonian encoding our solution. However, in the cases we tested, this approach also worked well for the initial time steps.

\subsection{Optimized scheduling time}

When we have no knowledge of the energetic landscape of an annealing evolution, the best scheduling time we can set is a simple linear function $s(t) = t/T, \; t\in[0,T]$, where $T$ is the total annealing time. In Qibo, this is encoded as $s(t) = t$, and the $T$ is managed internally. Thus, the boundary conditions become $s(0)=0$ and $s(1)=1$.

However, we have seen DMRG's ability to simulate the annealing run for a discrete set of time steps, which can help us determine a better scheduling function. We seek a function $s(t)$ that meets the following criteria:
\begin{itemize}
    \item $s(t)$ should dictate a faster evolution at time steps when the gap is larger, as there is less risk of an energy jump from the ground state to the first excited state (remember Section \ref{sec: AQC intro}).
    \item It should satisfy the boundary conditions $s(0)=0$ and $s(1)=1$.
\end{itemize}

\begin{figure}[htb]
    \centering
    \includegraphics[width= 0.7\columnwidth]{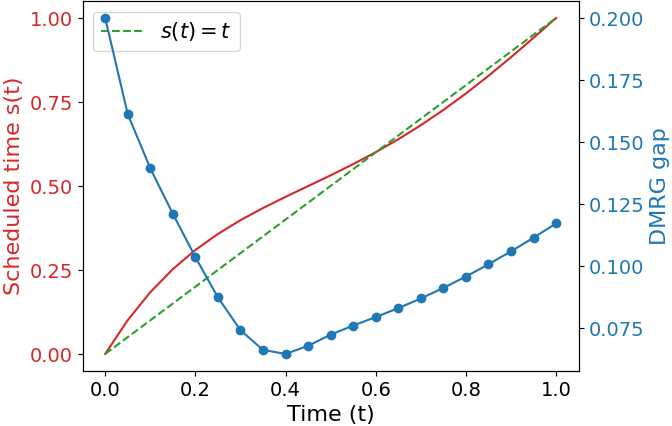}
    \caption{The plot displays the computed schedule time function for a QKP instance with 5 items. The algorithm first calculates the gaps for 20 equidistant time steps of the simulated annealing run using the DMRG algorithm. Subsequently, the scheduled time is computed to ensure faster evolution when the gap is larger, as the risk of an energy jump to the first excited state is lower. Both curves are shown in the same plot to illustrate how the gap influences the scheduling velocity.}
    \label{fig: schedule_time}
\end{figure}

Let's explain the steps we followed to compute our custom $s(t)$. Given a discrete set of gaps $\{ g_0, \ldots, g_N\}$ computed at equidistant time steps of the annealing run, we first determine a measure of the desired evolution velocity
\begin{align*}
    & v(\tilde{t}) =  \frac{g_{\tilde{t}} - g_{min} + \epsilon}{g_{max} - g_{min}} &&  \tilde{t} = 0, \ldots, N \; \; ,
\end{align*}
where $g_{min} = \min_i g_i$ and $g_{max} = \max_i g_i$ are the minimum and maximum gap found, and $\epsilon > 0$ is a parameter that ensures the velocity is not zero when $g_{\tilde{t}} = g_{min}$. This velocity satisfies
\begin{align*}
    & v(\tilde{t}) = \frac{\epsilon}{g_{max} - g_{min}} \approx 0 && \text{ when } g_{\tilde{t}} = g_{min}  \\
    \\
    & v(\tilde{t}) = \frac{g_{max} - g_{min} + \epsilon}{g_{max} - g_{min}}>1 && \text{ when } g_{\tilde{t}} = g_{max} \; \; .
\end{align*}
This behavior is what we expect, as the evolution proceeds faster when the gap is larger. We need to integrate this velocity to get a scheduling time function whose derivative matches this velocity. However, since our velocity is defined for a discrete set of points $\tilde{t}$, we need a continuous function for integration. To address this, we fit a polynomial to the discrete velocity values, and once we have the polynomial, we can integrate it
$$
\tilde{s}(t) = \int p(t) \, dt \; \text{ where $p(t)$ is the polynomial that best fits $v(\tilde{t})$  .}
$$
Finally, to ensure it satisfies the boundary conditions $s(0)=0$ and $s(1)=1$ we must do the following shift and scaling
$$
s(t) = \frac{\tilde{s}(t) - \tilde{s}(0)}{\tilde{s}(1) - \tilde{s}(0)} \; \; .
$$
You can check the result in Fig. \ref{fig: schedule_time}. Unfortunately, despite the scheduling appearing correct, we have not achieved significantly better results compared to the regular scheduling $s(t) = t$. We did not have time to delve deeper into this issue, but we believe exploring this in future work would be worthwhile.

\section{Large instances of the problem}

We do not want to conclude this thesis without exploring the limits of our work in terms of size. Unfortunately, we do not have access to a quantum annealer with a significant number of qubits, and Qibo simulations could not handle instances with more than approximately 10 qubits on our laptop.

Regarding DMRG, when tested with bigger instances, we observed that it was getting stuck in local minima: there was convergence, but far from the ground state. In future work, we would like to try to solve this issue, with strategies such as optimizing more than two sites in a single optimization step, or using other TN structures instead of MPS and MPO.

This leaves us with the simulated annealing solver provided by D-Wave in their Python library called Neal \cite{neal}. However, we needed a different solver to validate the correctness of the results. Initially, we implemented a brute-force approach for optimal results, but it was ineffective for big instances of the problem, so we needed a more efficient implementation. We opted for a dynamic programming algorithm (DP from now on), as it is the standard approach when solving a classical KP. Unfortunately, for the QKP achieving the optimal solution is not guaranteed. If you are curious about why this happens, refer to Appendix \ref{sec: DP algorithm}, where you will also find our implementation.

\begin{figure}[htb]
    \centering
    \includegraphics[width= 0.9\columnwidth]{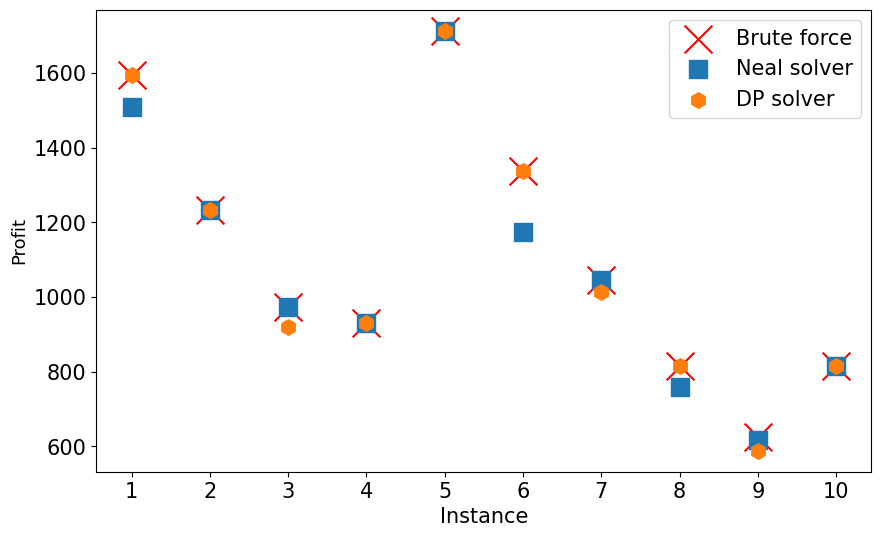}
    \caption{Solutions for random instances of the QKP involving 20 items and a maximum capacity of 100, where the values and weights of the items are randomly selected integers between 1 and 100.}
    \label{fig: results_bf}
\end{figure}

\begin{figure}[htb]
    \centering
    \includegraphics[width= 0.8\columnwidth]{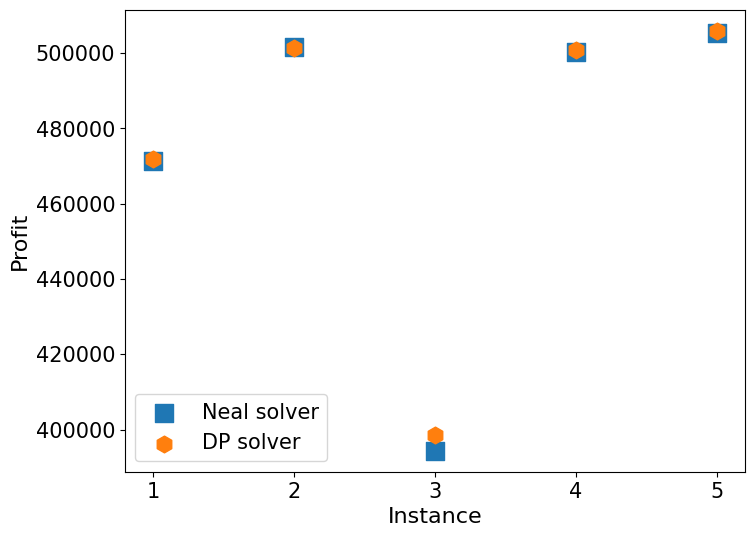}
    \caption{Solutions for random instances of the QKP involving 1000 items and a maximum capacity of 1000, where the values and weights of the items are randomly selected integers between 1 and 100.}
    \label{fig: results_large}
\end{figure}

To test the quality of the results we first tested small enough instances to be solved with brute force. Specifically, we executed random instances of the QKP with 20 items and a maximum capacity of 100, and we observed how the Neal and DP solvers have very similar performance, both reaching the optimal or very close to optimal solutions, as you can see in Fig. \ref{fig: results_bf}. After this quality test, we tested larger instances with 1000 items and 1000 weight capacity. Again, the two solvers yield very similar solutions, as you can appreciate in Fig. \ref{fig: results_large}.

Finally, if someone has to decide which solver is best for their use case, they should take into account:

\begin{itemize}
    \item The simulated annealing solver takes more time to compute than the DP solver for the instances we tested. However, ideally one would like to have a quantum annealer available and that would be a totally different scenario.
    \item The DP approach has a time complexity of roughly $O(W\cdot N^2)$ (see Appendix \ref{sec: DP algorithm}) so it depends on the maximum weight capacity. On the other hand, despite studying the time complexity of an annealing solver is not that straightforward, we can say that the QUBO formulation we are using with the unbalanced penalization technique does not depend on the weight capacity, which implies that increasing $W$ has no impact in the number of variables needed. 
    \item The annealing approach is minimizing energy instead of maximizing profit. In all the instances we could test with brute force the Neal solver was reaching the state with minimum energy, but the penalizations included in the QUBO formulation deform the solution space as we saw in Fig. \ref{fig: unbalanced penalization}, leading to suboptimal solutions. This is the price we pay to impose constraints in a QUBO formulation without introducing additional variables, necessitating extra care in fine-tuning the penalty parameters (the $\lambda$ value in Equation \ref{eq: QUBO cost}).
\end{itemize}

%% file: chapters/Conclusion.tex
\chapter{Conclusion and acknowledgements}

This thesis has ventured into the captivating intersection of quantum computing, tensor networks, and optimization problems. It has become clear that tensor networks are more than just a numerical tool; they represent a useful language for physics, mathematics, and computer science. While tensor networks embody the present capabilities, quantum computers hold the promise of a revolutionary future.

The technical contributions of this thesis are noteworthy, ranging from simplifying complex concepts for those without a physics background to reproducing and exploring state-of-the-art methodologies, such as creating Matrix Product Operators (MPOs) with finite automata and developing an annealing schedule based on estimations of the minimum gap using the Density Matrix Renormalization Group (DMRG) algorithm. Each chapter builds upon the previous, offering a comprehensive understanding of these cutting-edge technologies. Key takeaways include:

\begin{itemize}
    \item The versatility of Quadratic Unconstrained Binary Optimization (QUBO) formulation within quantum computing.
    \item The substantial capacity of the DMRG algorithm in simulating adiabatic evolution, potentially replacing exact diagonalization in Adiabatic Quantum Computing (AQC) studies.
    \item The clarity and visual representation provided by the automata framework in describing MPOs, simplifying the creation process.
    \item There is plenty of room for mathematicians and computer scientists in the context of quantum computing, as we all speak the same language: the language of mathematics.
    \item Classical approaches, like dynamic programming, should not be overlooked, as they can yield remarkably acceptable solutions with significantly less effort. The best solution is not always the optimal one; one should also consider the resources needed, including time of development and computational resources available.
\end{itemize}

While we have covered a broad spectrum of concepts, there remains much potential for further exploration. An additional focus on fine-tuning the numerous parameters of these algorithms could significantly impact the results obtained. Moreover, we were constrained by the limited computing power of personal computers. Access to a computing cluster or a real quantum annealer could have provided more valuable insights.

\subsubsection{Acknowledgements}

I would like to express my sincerest gratitude to all of my colleagues at \textbf{Qilimanjaro Quantum Tech}. You have been great at teaching me everything I know about quantum computing and making my time here so enjoyable. Andr{\'e}s, Jordi, Arnau, Ameer, Josep, Jose, Adri{\`a}, Pau, Ana, and Jan; your knowledge and support have been amazing, and I have learned so much from each of you. I want also to thank Marta P. Estarellas and Albert Solana for making this collaboration possible. I wish you all the best of luck with the company and its exciting future.

I also want to extend my heartfelt thanks to all the professors I encountered throughout my studies at the \textbf{University of Barcelona}. There are so many of you that I cannot mention you all, but please know that each of you has made a significant impact on my academic journey. A special thanks to my thesis directors, Nahuel and Luis, for their exceptional support and mentorship. Thank you for inspiring and challenging me every step of the way, I could not be happier with my choice to pursue this double degree.

Thank you for everything!